\documentclass[reprint,superscriptaddress,preprintnumbers,amsmath,amssymb,aps,pra]{revtex4-1}

\usepackage{graphicx}% Include figure files
\usepackage{dcolumn}% Align table columns on decimal point
\usepackage{bm}% bold math
\usepackage{siunitx}
\usepackage{color}

\newcommand{\SNR}{\text{SNR }}
\newcommand{\chii}{\chi_{\rm I}}
\newcommand{\chir}{\chi_{\rm R}}

\newcommand{\figref}[1]{Fig.\ \ref{#1}}

\newcommand{\eg}{\emph{e.g.\@}}
\newcommand{\ie}{\emph{i.e.\@}}

\newcommand{\eps}{\varepsilon}
\newcommand{\la}{\langle}
\newcommand{\ra}{\rangle}

\newcommand{\ls}{\left[ }
\newcommand{\rs}{\right] }
\newcommand{\lc}{\left\{ }
\newcommand{\rc}{\right\} }
\newcommand{\lr}{\left( }
\newcommand{\rr}{\right) }
\newcommand{\lb}{\left| }
\newcommand{\rb}{\right| }
\newcommand{\ld}{\left. }
\newcommand{\rd}{\right. }
\newcommand{\sv}[1]{\begin{bmatrix} #1 \end{bmatrix} }
\newcommand{\diff}{{\rm d}}
\newcommand{\vq}{\boldsymbol{q}}
\newcommand{\Op}[1]{\mathsf #1}
\newcommand{\oM}{ \Op{M} }

\begin{document}

\title{Full-field cavity enhanced microscopy techniques}
\author{Stefan Nimmrichter}
\affiliation{Centre for Quantum Technologies, National University of Singapore, 3 Science Drive 2, Singapore 117543, Singapore}
\author{Chi-Fang Chen}
\affiliation{Physics Department, Stanford University, 382 Via Pueblo Mall, Stanford, California 94305, USA}
\author{Brannon B. Klopfer}
\affiliation{Physics Department, Stanford University, 382 Via Pueblo Mall, Stanford, California 94305, USA}
\author{Mark A. Kasevich}
\affiliation{Physics Department, Stanford University, 382 Via Pueblo Mall, Stanford, California 94305, USA}
\author{Thomas Juffmann}
\affiliation{Physics Department, Stanford University, 382 Via Pueblo Mall, Stanford, California 94305, USA}
\affiliation{Laboratoire Kastler Brossel, ENS-PSL Research University, CNRS, 24 rue Lhomond, 75005 Paris, France}

%\email{cqtsn@nus.edu.sg}
%\date{\today}

\begin{abstract}
Quantum enhanced microscopy allows for measurements at high sensitivities and low damage. Recently, multi-pass microscopy was introduced as such a scheme, exploiting the sensitivity enhancement offered by multiple photon-sample interactions. Here we theoretically and numerically compare three different contrast enhancing techniques that are all based on self-imaging cavities: CW cavity enhanced microscopy, cavity ring-down microscopy and multi-pass microscopy. We show that all three schemes can lead to sensitivities beyond the standard quantum limit. 
\end{abstract}

\maketitle

\section{\label{sec:level1}Introduction}

Cavity enhanced measurements are ubiquitous in science and technology. In microscopy, the offered sensitivity enhancement has for example been exploited in cavity scanning microscopy \cite{Mader2015a,Hummer2015Cavity-enhancedNanotubes}, in Tolansky interferometry \cite{tolansky1970multiple} and in multi-pass microscopy \cite{Juffmann2016a,Klopfer2016IterativeLight}. While the former represents a point scanning technique, in which a fiber based microcavity is scanned across a sample, the latter two offer a full field of view.
In Tolansky interferometry, cavity enhancement is achieved by placing a flat mirror at a slight angle on top of the sample, which also has to be highly reflective. The incoupled light bounces back and forth between the two mirrors and the interference between multiple reflected beams is highly sensitive to the distance between the specimen and the mirror. Using this simple technique metallic surface topographies are routinely characterized on the nm level. However, the angle in between the two mirrors leads to beam walk-off and therefore to a non local response. This reduces the achievable transverse resolution to a few wavelengths of the probe light \cite{Hall1974Multiple-BeamInterferometry}. 
This can be avoided if the sample is placed in a self-imaging cavity \cite{Arnaud1969a,Gigan2005a}, as done in multi-pass microscopy \cite{Juffmann2016a}, a geometry that allows for 2D imaging and that is applicable to a wider range of samples, as long as photon loss is small. 

Here we analyze such cavity enhanced measurements based on self-imaging cavities and differentiate between three different regimes: 
The continuous wave scheme (CW), in which a continuous beam of light is in- and outcoupled into the self-imaging cavity via one of its end mirrors. 
The ring-down scheme (RD), in which a pulse of light is incoupled into the self-imaging cavity and a fraction of it is outcoupled every time the pulse interacts with one of the semi-transparent end mirrors of the self-imaging cavity. The detection can either be done in a time-resolved way, in which the number of interactions is recorded for each detected photon, or in a time-integrating way. 
The multi-pass scheme (MP) \cite{Juffmann2016a}, in which a pulse of light is incoupled into the self-imaging cavity and interacts with the specimen exactly $m$ times before it is outcoupled and detected. 
We first discuss these techniques analytically in the matrix optics formalism and derive expressions for the expected signal strength of bright-field (BF), dark-field (DF) and Zernike phase contrast (Znk) microscopy measurements (section \ref{sec:ray}). We then apply our findings to the cavity enhanced detection of mono- and few-atomic films of different materials, such as carbon and boron nitride (section \ref{sec:SNR}), and analyze the performance of each technique in terms of the achievable signal-to-noise ratio (SNR) per absorbed photon. We show that cavity enhanced microscopy techniques outperform classical microscopy techniques in these terms.
This agrees with results from the quantum measurement community \cite{Giovannetti2011a,Braun2017QuantumEntanglement}, where it has been shown that multi-passing represents a quantum optimal approach to phase measurements \cite{Giovannetti2006a} that allows overcoming the shot-noise limit and approaching the Heisenberg limit \cite{Luis2002a,Higgins2007a}. 
Besides the sensitivity enhancement for the detection of weak signals, cavity enhanced microscopy techniques will thus be of great interest for the study of photo-sensitive materials, for which a higher SNR cannot be achieved by using more probe photons. One example is live cell microscopy, where it has been shown that, even at visible wavelengths, long term observations or high-intensity microscopy techniques can cause cell death of a large fraction of a cell population \cite{Waldchen2015short}. MP microscopy has also been proposed for transmission electron microscopy \cite{Juffmann2016Multi-passMicroscopyb}, where sample damage sets bounds on the spatial resolution obtained for structural biology \cite{Glaeser2016}.

\section{\label{sec:ray}Theoretical model}

\begin{figure}
	\includegraphics[width=\columnwidth]{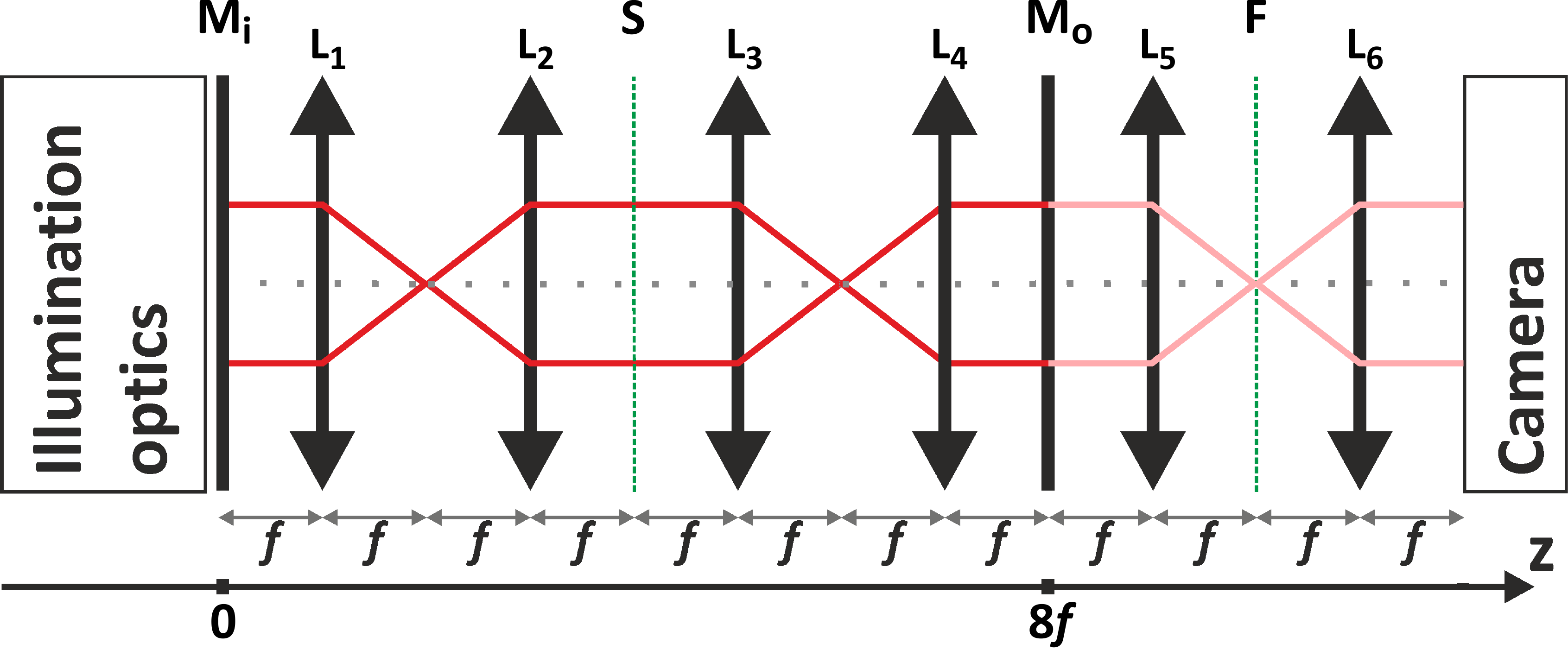}
	\caption{Setup for cavity enhanced microscopy. The self-imaging cavity consists of four lenses $L_{1\ldots 4}$ in between an incoupling and an outcoupling mirror ($M_i$ and $M_o$, respectively). The lenses are spaced such that, according to their respective focal lengths $f_{1\ldots 4}$, a microscope is formed on either side of the sample plane S. For simplicity we will restrict the following analysis to the case where $f_i=f$, resulting in unity magnification on either side of the sample. When the sample is illuminated from the left, a mirror-flipped image will be formed on $M_o$ and the reflected light will be re-imaged onto the sample, which is now illuminated with an image of itself. After multiple $m$ interactions, light is either actively or passively outcoupled through $M_o$ and imaged using the microscope to the right of $M_o$, where additional optics in the Fourier plane allow for dark field and phase imaging. \label{fig:s1}}
\end{figure}

A setup for cavity enhanced microscopy is shown in Fig.~\ref{fig:s1}. The following analysis will be restricted to the paraxial ray-optics regime, which is a good approximation in cases not affected by the diffraction limit. It will further be restricted to scalar fields. In this idealized scenario, the self-imaging cavity between the two mirrors at $z=0$ and $z=8f$ is comprised of infinite thin lenses for perfect imaging, two idealized beam splitters as a model for the semi transparent mirrors, and a thin refractive index profile representing the sample plane in the center of the arrangement at $z=4f$. Given an input light field coupled in from the left ($z=0$), the optical response of the sample will be encoded in the amplitude and phase of the field outcoupled through the right mirror ($z=8f$) after multiple cavity roundtrips and sample interactions. First the analytic expressions for the outcoupled field and the energy absorbed by the sample will be derived in section \ref{sec:toy} and section \ref{sec:damage}, respectively. In section \ref{sec:toyMeasurement}, a possible post-processing in a subsequent 4$f$ lens arrangement for Zernike and dark-field imaging \cite{Zernike1942,mertz2009introduction} will be discussed. 

\subsection{\label{sec:toy}The $8f$ imaging cavity}

Let us start by modelling the self-imaging cavity 
and the effect on the light field as it bounces between the mirrors and repeatedly interacts with the sample. 
The input field illuminating the first mirror is assumed to be a broad Gaussian mode with central frequency $\omega$, a waist $w$ greater than the sample dimensions, and a possibly time-dependent input power $P_{\rm in} (t)$, 
\begin{equation}
E_{\rm in}(x,y,t) = \sqrt{\frac{4P_{\rm in} (t)}{\pi c \eps_0 w^2}} e^{-(x^2+y^2)/w^2 - i\omega t}. \label{eq:Ein_gauss}
\end{equation}
We will study both a continuous wave scenario with time-independent input power and a pulsed scenario. In the latter case, we restrict our considerations to moderate pulse lengths: They can be short enough to prevent the fields of subsequent roundtrips from overlapping, while they are still sufficiently long to neglect the variation of the longitudinal wavelength, $\omega = ck$, for all frequency-dependent interaction processes. The longitudinal phase factor $\exp (ikz)$ is omitted as we also operate in the paraxial regime. 

Each mirror shall be treated as a simple beam splitter with amplitude reflectivity and transmissivity parameters $r_{1,2}=-\sqrt{R_{1,2}}$ and $t_{1,2}$, where $T_{1,2}=|t_{1,2}|^2$ and $R_{1,2} \gg T_{1,2}$. The propagation of the field through the lenses between mirror and sample planes is described by an ideal (infinite-aperture) $4f$ transformation \cite{mertz2009introduction} that produces the inverted image,
\begin{equation}
E(x,y,t) \to - E \lr -x,-y, t-\frac{4f}{c} \rr \label{eq:invert4f}
\end{equation}
A typical sample inserted into the cavity would be given by a thin layer on top of a transparent carrier plate of known refractive index $n_g$ and thickness $d_g \ll f$. Absorption and diffraction within the carrier are neglected. The sample layer that is to be detected shall be described by a two-dimensional refractive index profile $n(x,y)$ of thickness $d$. The real and the imaginary part of $n(x,y)$ will be imprinted in the detection signal of the microscope, \ie{} the phase and the amplitude of a probe light field. Unlike in single-pass imaging, which `sees' only the transmission profile of the sample (and holder), the present two-mirror multi-pass scheme requires us to take also the reflectivity of the sample into account. In the macroscopic limit of perfectly resolved sample structure, the transmission and reflection can be obtained by solving the boundary conditions at the interfaces of sample layer and holder material per `pixel' $(x,y)$ on the sample plane. 
For readability, we will omit the argument in the following and abbreviate $n_s = n(x,y)$, keeping in mind that all expressions are defined per pixel.

The carrier glass slab is characterized by the transmission and reflection coefficients \cite{Orfanidis2002}
\begin{eqnarray}
t_g &=& \frac{4n_g e^{i(n_g-1)kd_g}}{(n_g + 1)^2 - (n_g - 1)^2e^{2in_g kd_g}}, \nonumber \\
r_g &=& \frac{(n_g^2 - 1) \lr e^{2in_g k d_g} - 1 \rr}{(n_g + 1)^2 - (n_g - 1)^2e^{2in_g kd_g}}. \label{eq:rt_glass}
\end{eqnarray}
For a clean signature of the substrate, the carrier can be made perfectly transmissive (non-reflective) by choosing its optical thickness to be a half-multiple wavelength, $n_g k d_g=j\pi$, $t_g = (-)^j e^{-ikd_g}$, $r_g = 0 $.
In this case, we obtain relatively simple expressions for the transmission and reflection coefficients of the whole sample, 
\begin{eqnarray}
t_s &=& \frac{4 (-)^j n_s e^{i\lr n_s -1 \rr kd-ikd_g}}{\lr n_s+1\rr^{2}-\lr n_s-1\rr^{2} e^{2in_s kd} } , \nonumber \\
r_{s,L} &=& \frac{ \lr n_s^2 - 1 \rr e^{-2ikd} \lr e^{2i n_s kd} - 1 \rr}{\lr n_s+1\rr^{2}-\lr n_s-1\rr^{2} e^{2in_s kd} } \nonumber \\
r_{s,R} &=& e^{2i k (d - d_g)} r_{s,L}. \label{eq:rt_sample}
\end{eqnarray}
Notice the difference in the reflection of fields impinging on the substrate side ($L$) and on the back side ($R$). All reflected and transmitted field components are defined relative to the incident field on the sample plane, $z=4f$, which is set to be the interface between the substrate layer (to the left) and the carrier plate (right).

As the main application of the scheme is to enhance weak optical signatures, we focus here on optically thin substrate layers, $|n_s| k d \ll 1$. To lowest order, their optical response can be characterized by the susceptibility function 
\begin{equation}
\chi = \chir  + i\chii  = \frac{n_s^2-1}{2} k d, \label{eq:susc}
\end{equation}
where the real part represents the sample-induced phase shift and the imaginary part the extinction of the incident field amplitude. 
The above sample coefficients are approximated by
\begin{eqnarray}
t_s &\approx& (-)^j e^{-i k d_g} \lr 1 + i \chi \rr \approx e^{ij\pi -ikd_g +i\chi} , \nonumber \\ 
r_{s,L} &\approx& i \chi , \quad 
r_{s,R} \approx i e^{-2ikd_g} \chi. \label{eq:sampleWS}
\end{eqnarray}
It turns out that the validity of these linearized expressions is limited in practice when it comes to the quantitative analysis of multi-pass imaging of thin films (see Sect.~\ref{sec:SNR}). Nevertheless it can serve as a qualitative estimate for the signal enhancement in multi-pass imaging, and we shall occasionally refer to this as the weak-sample (WS) scenario later. 

In order to describe the transformation of an arbitrary input pulse at the $8f$-cavity-sample system into a (possibly overlapping) sequence of output pulses, we can make use of the matrix optics formalism \cite{Siegman1986Lasers}. Given the light fields $E_{L\rightarrow}$, $E_{R\leftarrow}$ impinging on the sample plane from the left and right (with the arrows marking the propagation direction), the sample interaction is described by a linear map for each sample pixel,
\begin{equation}
\sv{E_{L\leftarrow} (t) \\ E_{R\rightarrow} (t)} = \oM_s \sv{E_{L\rightarrow} (t) \\ E_{R\leftarrow} (t)}, \quad 
\oM_s = \sv{ r_{s,L} & t_s \\ t_s & r_{s,R} }. \label{eq:Ms}
\end{equation}
Here, we assume that the sample interaction and the cavity properties are approximately constant over the pulse spectrum (\ie{} determined by their values at the central frequency of the pulse). The passage back and forth through the $4f$ lens systems and reflection at the two outer mirrors can be expressed by $ E_{L\rightarrow} (t+8f/c) = r_1 E_{L\leftarrow} (t)$ and $E_{R\leftarrow} (t+8f/c) = r_2 E_{R\rightarrow} (t)$. We arrive at the following transformation matrix,
\begin{equation}
\sv{E_{L\rightarrow} (t+\tfrac{8f}{c}) \\ E_{R\leftarrow} (t+\tfrac{8f}{c})} = \oM \sv{E_{L\rightarrow} (t) \\ E_{R\leftarrow} (t)}, \, \, \oM = \sv{ r_1 r_{s,L} & r_1 t_s \\ r_2 t_s & r_2 r_{s,R} },
\end{equation}
for each sample pass followed by a half roundtrip. At this point, we shall introduce the eigenvalues of this matrix for later use,
\begin{equation}
\lambda_{\pm} = \frac{r_1 r_{s,L} + r_2 r_{s,R} \pm \sqrt{ \lr r_1 r_{s,L} - r_2 r_{s,R} \rr^2 + 4r_1 r_2 t_s^2 } }{2}. \label{eq:M_eig}
\end{equation}
Given an incident pulse $E_{\rm in} (t)$ that is initially coupled in through the left mirror ($z=0$), the forward-running pulse amplitude on the sample plane ($z=4f$) after $m\geq 1$ passes through the sample reads as
\begin{eqnarray}
E^{(m)}_{R\rightarrow} (t) &=& -t_1 E_{\rm in} \ls t - (2m-1) \tfrac{4f}{c}\rs \sv{0 \\ 1} \cdot \oM_s \oM^{m-1} \sv{ 1 \\ 0} \nonumber \\
&=& - \frac{\lambda_{+}^m - \lambda_{-}^m }{\lambda_{+} - \lambda_{-} } t_1 t_s E_{\rm in} \ls t - (2m-1) \tfrac{4f}{c}\rs. 
\end{eqnarray}
Note that the coordinate inversion by the first 4f transformation according to \eqref{eq:invert4f} leaves the gaussian input field \eqref{eq:Ein_gauss} invariant. 
The outcoupled train of pulses at the second mirror ($z=8f$) is simply obtained by taking the sum over $m$ and multiplying with the transmission of the second mirror,
\begin{equation}
E_{\rm out}(t) = t_1 t_2 t_s \sum_{m=1}^{\infty} \frac{ \lambda_{+}^m - \lambda_{-}^m}{\lambda_{+} - \lambda_{-}}  E_{\rm in}\lr t-\tfrac{8m f}{c}\rr. \label{eq:EoutTime_ideal}
\end{equation}
Again, the output field is inverted with respect to the sample plane, \ie{} the sample pixel $(x,y)$ is imaged onto $(-x,-y)$. 

The input-output transformation can also be given in Fourier space, which for a fixed light frequency $\omega$ amounts to a stationary illumination, \ie{} infinite pulse length. Given the temporal Fourier transform $E_{\rm in} (\omega)$ of the input field \eqref{eq:Ein_gauss} the transmitted output field becomes
\begin{eqnarray}
E_{\rm out} (\omega) = \frac{t_1 t_2 t_s E_{\rm in}(\omega) e^{8ikf}}{ \lr 1 - e^{8ikf} \lambda_{+} \rr \lr 1 - e^{8ikf} \lambda_{-} \rr}.  \label{eq:Eout_ideal}
\end{eqnarray}
It follows either by carrying out the sum in \eqref{eq:EoutTime_ideal} in Fourier space, or by directly solving the combined boundary value problem at the sample plane and the mirrors. 
Additional losses in the cavity, \eg{} at the lenses, can be included by setting $R_{1,2} + T_{1,2} < 1$. 

Our main focus here are samples with weak optical response, which implies a low overall reflectivity, $|r_{s,(L,R)}|^2 \ll |t_{s}|^2 \lesssim 1$. 
However, the degree to which the sample reflectivity influences the multi-pass image depends also on the reflectivity of the sample holder and on the number of roundtrips. If the $8f$ cavity is of high finesse, \ie{} supports many roundtrips, multiple sample- or holder-reflected fields interfere and may have a significant impact on the cavity resonance and on the output field.

In the following, we distinguish two complementary regimes for cavity-enhanced microscopy by comparing the characteristic duration $\tau$ of the input pulse $P_{\rm in} (t)$ to the half round-trip time $8f/c$. A quasi-stationary frequency-domain description, Eq.~\eqref{eq:Eout_ideal}, applies in the continuous-wave (CW) limit $\tau\gg 8f/c$, whereas a time-domain treatment, Eq.~\eqref{eq:EoutTime_ideal}, of individual non-overlapping pulses is more suitable for $\tau < 8f/c$, \ie{} in the multi-pass (MP) and ring-down (RD) cases. 
The intensity of the outcoupled light is then either given by the interference of many field components or a sum of individual pulses.

The sample response in the output field can be made explicit in the WS limit \eqref{eq:sampleWS}. Using the approximation $1+x \approx e^x$ for $|x|\ll 1$, we obtain to lowest order
\begin{widetext}
\begin{eqnarray}
E_{\rm out} (\omega) &\approx & \frac{(-)^j t_{1}t_{2}E_{\rm in}(\omega) e^{8ikf-ikd_g + i \chi}}{1-r_{1}r_{2} e^{16ikf-2ikd_g+2i\chi} - i \chi e^{8ikf-ikd_g} \ls r_{1}e^{i k d_g} + r_{2}e^{-i k d_g}\rs }, \label{eq:Eout_weak_w} \\
E_{\rm out} (t) &\approx & (-)^j t_{1} t_{2} e^{-i k d_g + i \chi } \sum_{\ell=0}^{\infty} \lr r_{1}r_{2} e^{2i\chi - 2ikd_g} \rr^{\ell} \lc E_{\rm in} \ls t-\tfrac{8f}{c} (2\ell+1) \rs + i \ell \chi \frac{r_{1}e^{2i k d_g} + r_{2} }{r_{1}r_{2}} E_{\rm in} \ls t-\tfrac{16 \ell f}{c} \rs \rc. \label{eq:Eout_weak_t}
\end{eqnarray}
\end{widetext}
The terms $\chi$ and $kd_g$ are to be evaluated by their values at the mean pulse wave number $k$. The time-domain expression \eqref{eq:Eout_weak_t} splits into contributions associated to odd and to even numbers of sample interactions. The latter terms describe the light that is reflected at the sample, whereas the former correspond to one pass and $\ell$ additional full roundtrips in the resonator, \ie{} to $m=2\ell+1$ sample interactions, a total phase shift of $(2\ell +1) \chir $, and an extinction of $(2\ell + 1)\chii $ per pixel. The signal enhancement by the number of passes is the key feature of the studied MP imaging scheme, as we will discuss below. 

In the limit of stationary illumination, the mirror system acts like a resonator, and the signal enhancement is expected to scale with the cavity finesse, \ie{} the average number of photon round-trips, see Sect.\ref{sec:rayCW}.

\subsection{\label{sec:damage}Sample damage}
Apart from optical resolution, a key limitation for microscopy with sensitive biological samples is the damage induced by photon absorption. It sets the gauge for comparing the performance of multi-pass imaging and conventional single-pass microscopy: We can rank the performance of different microscopy schemes by the SNR of the (phase or absorption) images they produce at a fixed threshold value for the overall sample damage. 
In the following we will assume that the damage is proportional to the amount of energy that is absorbed by the sample. 

The net absorbed power per sample pixel is formally obtained by summing the inward-oriented Poynting vectors left and right of the sample plane, assuming that the sample holder is transparent. Here, this amounts to comparing the forward- and backward-running intensities,
\begin{equation}
I_{\rm abs}(t) = I_{L\rightarrow} (t) - I_{R\rightarrow} (t) + I_{R\leftarrow} (t) - I_{L\leftarrow} (t). \label{eq:Iabs}
\end{equation}
In the case of stationary illumination, this is directly proportional to the sample damage rate. The fields left and right of the sample follow by solving the boundary conditions and can be expressed in terms of the output field \eqref{eq:Eout_ideal} at pixel $(-x,-y)$. We arrive at a damage rate proportional to the cavity-enhanced output intensity,
\begin{eqnarray}
I_{\rm abs}(\omega) &=& \frac{I_{\rm out} (\omega) }{T_2} \ls  R_2 - 1 + \lb \frac{1 - e^{8ikf} r_2 r_{s,R} }{t_s} \rb^2 \rd \nonumber \\
&&- \ld \lb \frac{r_{s,L} + e^{8ikf} r_2 \lr t_s^{2}- r_{s,L}r_{s,R} \rr }{t_s } \rb^2 \rs.
\end{eqnarray}
For time-dependent input fields, the overall absorbed energy $Q_{\rm abs}$ per pixel (corresponding to $N_{\rm abs}=Q_{\rm abs}/ \hbar \omega $ absorbed photons) is obtained by integrating the intensity \eqref{eq:Iabs} over the interrogation time and the pixel area. 
We conveniently express the fields on both sides of the sample in matrix notation, using \eqref{eq:Ms} and 
\begin{eqnarray}
\!\!\!\! \sv{E_{L\rightarrow} (t) \\ E_{R\leftarrow} (t)}&=&  -t_1 \sum_{m=0}^\infty \oM^{m} \sv{E_{\rm in} \lr t - \tfrac{8mf}{c} - \tfrac{4f}{c} \rr \\ 0} \! ,
\end{eqnarray}
where each summand represents the field after $m$ sample interactions. A handy result is found in the case of non-overlapping roundtrip pulses. Given the temporal power profile $P_{\rm in}(t)$ of the input pulse with characteristic duration $\tau$ and a small pixel area in the center of the pulse profile, $A \ll w^2$, the input energy per pixel is
\begin{equation}
Q_{\rm in} = \frac{2A}{\pi w^2} \int_\tau \diff t \, P_{\rm in}(t) . \label{eq:Qin}
\end{equation}
Assuming also a constant (average) sample response over the size of each pixel, 
the absorbed energy per pixel accumulated after $m$ interactions reads as 
\begin{equation}
Q_{\rm abs}^{(m)} = T_1 Q_{\rm in} \sum_{n=0}^{m-1} \lc \lb \oM^{n} \sv{1 \\ 0} \rb^2 - \lb \oM_s \oM^{n} \sv{1 \\ 0} \rb^2 \rc. \label{eq:damage_m}
\end{equation}
In the WS limit \eqref{eq:sampleWS}, we find that the stationary absorption rate is proportional to the intra-cavity intensity times the absorption strength of the sample,
\begin{equation}
I_{\rm abs}(\omega) \approx 2 \chii \lb 1 + r_2 e^{8ikf-2ikd_g} \rb^2 \frac{I_{\rm out} (\omega)}{T_2} .\label{eq:IabsWS}
\end{equation}
For short pulses, a self-explanatory WS expression arises if $R_{1,2}=R$,
\begin{eqnarray}
Q_{\rm abs}^{(m)} &\approx& 2T_1 Q_{\rm in} \chii \ls \frac{1+R_2}{1-R_1 R_2} \rd \nonumber \\
&& \ld  - \frac{(r_1 r_2)^m}{2r_1} \lr \frac{r_1+r_2}{1-r_1 r_2} + (-)^m \frac{r_1 - r_2}{1+r_1 r_2} \rr \rs \nonumber \\ 
&\to& 2 T \frac{1-R^m}{1-R} Q_{\rm in} \chii.
\label{eq:damage_mWS}
\end{eqnarray}

\subsection{Amplitude and phase measurements \label{sec:toyMeasurement}}

In this ideal scenario of perfect resolution (neither limited by finite apertures nor by a finite pixel size in the detector) the output field carries full information about the local amplitude and phase modulation for each point on the sample plane. The measurement sensitivity would be limited only by shot noise. Depending on what information is to be extracted, we distinguish three detection schemes: a direct measurement of the local output intensity to image the light extinction profile of the sample (BF), a background-free dark-field measurement of the diffraction profile (DF), and a Zernike phase measurement \cite{Zernike1942} (Znk). 
In the short-pulsed regime, the detection signals are sequences of pulses arranged according to the number $m$ of interactions with the sample (\ie{} half roundtrips through the $8f$ imaging cavity). We shall refer to their individual per-pixel energies as $Q^{(m)}_{\rm BF,DF,Znk\pm}$.
In a multi-pass scheme where the $m$th pulse is outcoupled by a specific triggered mechanism, and not through the second cavity mirror, the factor $T_2$ in $Q^{(m)}_{\rm BF,DF,Znk\pm}$ must be replaced by the transmission efficiency of the outcoupler.

In the bright-field case (BF), the time-resolved detection signal will be determined by the absolute square of the field \eqref{eq:EoutTime_ideal}. 
For non-overlapping short roundtrip pulses, the square of the sum of the fields reduces to a sum of squares, and we obtain the bright-field signal
\begin{eqnarray}
Q_{\rm BF}^{(m)} &=& T_1 T_2 Q_{\rm in} \lb t_{s} \frac{\lambda_{+}^m - \lambda_{-}^m}{\lambda_{+} -\lambda_{-}} \rb^2 - Q_{\rm ref}^{(m)}, \label{eq:QbrightField}
\end{eqnarray}
once again evaluated at the mirrored image pixel of the sample.
This rather featureless expression exhibits a dichotomic behavior between odd and even numbers $m$ of sample interactions. The input light and the most significant sample response appears in the transmission signal after full cavity roundtrips, \ie{} odd $m=2\ell+1$. The signal after an even number of interactions implies at least one reflection at the sample (or sample holder) and is thus of higher order in its optical response. 

In BF microscopy of weak samples, most of the output light is just the transmitted input beam distributed over many roundtrips, with a small sample-induced modulation. We thus define the actual sample signal in each pulse relative to a reference $Q_{\rm ref}^{(m)}$, which could be another spot on the detection plane with a different sample profile (in a differential measurement), or an empty reference pixel. In the latter case, we obtain the reference signal from the output field of the 8f imaging cavity and an empty sample plate, assuming homogeneous illumination. It has the same form as \eqref{eq:EoutTime_ideal}, but with the reflection and transmission coefficients \eqref{eq:rt_glass} of the empty glass plate in place of the sample terms. The eigenvalues of the corresponding round-trip matrix are
\begin{equation}
\Lambda_{\pm} = \frac{\lr r_1 + r_2 \rr r_g \pm \sqrt{ \lr r_1 - r_2 \rr^2 r_g^2 + 4r_1 r_2 t_g^2 } }{2} , \label{eq:M_eig_glass} 
\end{equation}
They simplify to $\pm \sqrt{r_1 r_2} e^{-ikd_g}$ in the non-reflective case, $n_g k d_g = j\pi$. We arrive at
\begin{equation}
Q_{\rm ref}^{(m)} = T_1 T_2 Q_{\rm in} \lb t_{g} \frac{\Lambda_{+}^m -\Lambda_{-}^m}{\Lambda_{+} -\Lambda_{-}} \rb^2. \label{eq:Qref}
\end{equation}
For non-reflective holders it is nonzero only after odd multiples $m=2\ell+1$, where it simplifies to $T_1 T_2  Q_{\rm in} (R_1 R_2)^\ell$.
Once again, we get a clearer picture in the WS limit \eqref{eq:sampleWS}. Expanding the eigenvalues \eqref{eq:M_eig} to lowest non-vanishing order in $\chi$ and integrating over the pulse duration, we obtain the BF signal 
\begin{equation}
Q_{\rm BF}^{(m=2\ell+1)} \approx -2 m \chii Q_{\rm ref}^{(m)} . \label{eq:BF_pulse_odd} 
\end{equation}
for odd sample interactions, \ie{} full roundtrips. The result is negative due to the accumulated extinction of the input pulse at the sample, while the phase response does not enter this first order expression. In fact, the validity of the approximation is restricted to not too many roundtrips and to samples with significant absorption, $m|\chi| \ll 2\pi$ and $\chir^2 \ll \chii$.
The signal in between full roundtrips at even $m=2\ell$ is comprised of light reflected at the sample and is therefore of second order,
\begin{equation}
Q_{\rm BF}^{(m=2\ell)} \approx \lb m \chi \rb^2 \frac{\lb r_1 e^{2ikd_g} + r_2 \rb^2}{4} Q_{\rm ref}^{(m-1)}. \label{eq:BF_pulse_even}
\end{equation}

In dark-field and Zernike phase imaging, the outcoupled field \eqref{eq:EoutTime_ideal} passes another $4f$ configuration before detection, hitting either a small absorber (DF) or a phase plate (Znk) in the Fourier plane at the distance $2f$ behind the exit mirror.
The $2f$ transformation of a paraxial field amplitude yields the spatial Fourier transform \cite{Saleh1991,mertz2009introduction}, 
\begin{eqnarray}
E_{\rm out} (x,y,t) &\to& 
\frac{k e^{2ikf}}{2\pi i f} \tilde{E}_{\rm out} \lr \tfrac{k}{f}x, \tfrac{k}{f}y, t-\tfrac{2f}{c} \rr . \label{eq:2f_paraxial}
\end{eqnarray}
Being subject to a thin absorbing or phase-shifting plate, the field is then multiplied by a transmission function $[1-b(x,y)]$, modulating its amplitude or phase where $b(x,y)\neq 0$. This is followed by another $2f$ transform leading to the detection field, 
\begin{eqnarray}
E_{\rm det} (x,y,t) &=&  -e^{4ikf} \ls E_{\rm out} \lr -x, -y, t-\tfrac{4f}{c} \rr \rd \nonumber \\
&& \ld  - E_{\rm b} \lr -x, -y, t-\tfrac{4f}{c} \rr  \rs, \label{eq:4fplate_paraxial} \\
E_{\rm b} ( x, y, t) &=& \int \frac{k^2 \diff x' \diff y'}{(2\pi f)^2} \tilde{b} \lr \tfrac{k}{f} x', \tfrac{k}{f}y'\rr \nonumber \\
&&\times E_{\rm out} \lr x'+x, y'+y, t\rr . \label{eq:4fplate_block}
\end{eqnarray}
The dark-field image of a homogeneously illuminated sample structure is obtained by blocking the undiffracted forward component from the outcoupled field. This can be realized here by placing an absorbing element in the origin of the Fourier plane (see Fig.~\ref{fig:s1}), \eg{} a circular obstacle with radius $\varrho > f/kw$, 
so that 
$\tilde{b} (\vq) = J_1 \lr q \varrho \rr 2\pi \varrho /q$.

If the relevant sample size is much smaller than the Gaussian waist $w$ of the incident probe field \eqref{eq:Ein_gauss}, we can choose an absorber size $\varrho$ that blocks only the undiffracted beam and lets almost all the diffracted light pass. The blocked field \eqref{eq:4fplate_block} is then approximately given by the output field of the 8f imaging cavity with an empty sample plate. Note that the sample pixel is now imaged onto the same pixel on the detection plane.

In the short-pulse limit, the dark-field detection signal becomes
\begin{eqnarray}
Q_{\rm DF}^{(m)} &=& T_1 T_2 Q_{\rm in} \lb t_{s} \frac{\lambda_{+}^m - \lambda_{-}^m }{\lambda_{+} -\lambda_{-} } - t_g \frac{\Lambda_{+}^m-\Lambda_{-}^m}{\Lambda_{+}-\Lambda_{-}}\rb^2 .  \label{eq:IdarkField}
\end{eqnarray}
The output pulses associated to even and odd sample interactions are now of the same magnitude, and there is no need to subtract another reference term. In the WS limit \eqref{eq:sampleWS}, the even orders are identical to \eqref{eq:BF_pulse_even} before, and the odd ones are also of second order in the weak sample response, 
\begin{equation}
Q_{\rm DF}^{(m=2\ell+1)} \approx \lb m \chi \rb^2 Q_{\rm ref}^{(m)} . \label{eq:IdarkFieldWS}
\end{equation}

For the Zernike phase contrast method, the opaque plate in the Fourier plane behind the exit mirror is replaced by a phase plate that shifts the undiffracted background field component by $\pm \pi/2$ \cite{Zernike1942}. The field arriving at the detector in the Zernike scheme can be understood as a superposition of the background-free dark-field signal and the $\pi/2$-shifted undiffracted field without sample. Depending on the sign of the phase shift the technique is referred to as negative (Znk-) or positive (Znk+) phase contrast microscopy.
We obtain the signal from the DF case by inserting a complex prefactor, 
$b(x,y) \to \lr 1 \mp i \rr b(x,y)$. 
Repeating the above approximation steps then yields
\begin{eqnarray}
Q_{\rm Znk\pm}^{(m)} &=& T_1 T_2 Q_{\rm in} \lb t_{s}\frac{\lambda_{+}^m - \lambda_{-}^m }{\lambda_{+} -\lambda_{-} } - (1\mp i) t_g \frac{\Lambda_{+}^m-\Lambda_{-}^m}{\Lambda_{+}-\Lambda_{-}}\rb^2 \nonumber \\
&& - Q_{\rm ref}^{(m)}, \label{eq:IZernike}
\end{eqnarray}
Once again, we subtract the bright offset from the actual sample response, because, contrary to the DF case, the phase plate does not remove the reference signal \eqref{eq:Qref}.
The Zernike configuration can provide strong signals even for weak phase shifts of optically thin, transparent samples, as the phase response now appears in first order after full roundtrips. The WS limit yields 
\begin{equation}
Q_{\rm Znk\pm}^{(m=2\ell+1)} \approx \pm 2 m \chir Q_{\rm ref}^{(m)} . \label{eq:IZernikeWS}
\end{equation}
It has the same form as the BF signal \eqref{eq:BF_pulse_odd}, but with $\mp \chir$ instead of $\chii$. 

\subsection{Enhanced phase estimation by multi-passing}

We have shown that the phase or extinction signature of weak optical samples is generally enhanced linearly (in BF and Znk schemes) or quadratically (DF) by the number $m$ 
of times a probe field interacts with the specimen in the imaging cavity. This gain in measurement sensitivity with respect to the shot noise-limited accuracy of a single-pass microscope becomes apparent if we view the WS imaging as a parameter estimation problem.

In the absence of extinction losses and sample holder, a WS imprints the phase $\chi = \chir$ onto the coherent probe light upon each interaction. This phase can be estimated in the Znk$+$ scheme, where the purpose of the phase plate in the outcoupling stage is to interfere the phase-shifted component of the probe field with the unshifted one. The MP scheme implements the sequential application of the phase shift to one of the two interfered components. We then estimate the phase after $m$ passes by means of the difference between the detected photon numbers of the sample pixel and an empty reference pixel. Using simple error propagation of the respective shot noise, we get a mean estimate and error \cite{Giovannetti2004a,Giovannetti2006a} 
\begin{equation}
\chi_{\rm est} \approx \frac{N-N_{\rm ref}}{2m N_{\rm ref}}, \quad \delta \chi \approx \frac{\sqrt{N+N_{\rm ref}}}{2m N_{\rm ref}},
\end{equation}
with $N$ and $N_{\rm ref}$ the mean photon numbers of sample and reference pixel, respectively. Given that the latter are of about the same magnitude, we find $\delta \chi \sim 1/\sqrt{2N_{\rm ref}}m$. 

If sample damage is an issue, one can adjust the input intensity for a fixed number of photon-sample interactions, \ie{} constant $N_{\rm ref} m$. In this case, which will be studied in detail below, the error scales like $1/\sqrt{m}$. For an equivalent CW or RD detection scheme, the same proportionality holds with $2\la \ell \ra \gg 1$ instead of $m$.

\section{\label{sec:SNR} Signal to noise at constant damage}

We will now compare the various imaging modalities in terms of signal to noise at constant damage. As an illustrative example we will discuss the use of cavity enhanced microscopy for the characterization of ultrathin films of carbon and boron nitride (BN). Density functional theory calculations yield an index of refraction of graphene of $2.71 + 1.41i$ \cite{Cheon2014HowConstraints}. At this wavelength BN has a refractive index of $1.8$ \cite{Golla2013OpticalFlakes}, where the imaginary part is negligible due to the large bandgap of 5eV \cite{Ajayan2016Two-dimensionalMaterials}. The susceptibility of the two materials can be obtained from Eq.~\eqref{eq:susc}. The thickness of a monolayer of graphene and BN is \SI{3.35}{\angstrom} and \SI{3.33}{\angstrom} \cite{Golla2013OpticalFlakes}, respectively. 

In traditional microscopy these samples provide very low contrast and their detectability depends strongly on the thickness of the substrate and the probing wavelength \cite{Blake2007MakingVisible,Gorbachev2011HuntingSignatures}. 
Already now, several optical techniques are being used to characterize thin film growth \cite{Frey2015MeasurementsProcess}, and interferometric multibeam interference schemes, such as Tolansky interferometry \cite{Tolansky1951TheInterferometry}, are used to enhance the measurement sensitivity \cite{Raz1996AInterferometer}. The self-imaging capabilities of the cavity enhanced microscopy techniques discussed here offer the advantage that spatial film thickness variations can be detected locally (and in principle at diffraction-limited resolution).

So let us assume a sample with a substrate layer spatially varying in thickness or material. In order to detect such variations, we shall select two areas of the substrate that differ in their optical responses and image them onto two different pixels, $(x_1,y_1)$ and $(x_2,y_2)$. Depending on the imaging scheme, $N_{1,2}$ photons will be detected on the two pixels, respectively (where the photon number is given by the pulse energy divided by $\hbar \omega$). 
In a differential measurement, and assuming shot noise in the photodetector, the SNR of detecting the variation will then be $\SNR=|N_1-N_2|/\sqrt[]{N_1+N_2}$. We will evaluate and compare it at a fixed damage level for continuous-wave, ring-down and multi-pass microscopy, using the various imaging modes discussed in the previous chapter. 
For simplicity, we focus on the case studied in the previous section where the second pixel corresponds to an empty reference area, $\chi(x_2,y_2)=0$. At uniform illumination, this is equivalent to imaging a single pixel with and without the substrate layer. 
We assume uniform illumination of the relevant pixels with the same incident light energy $T_1 Q_{\rm in}$. 

Due to the low sample absorption and high number of roundtrips considered, the WS approximation will not always be valid; we use it to discuss the qualitative scaling of the signal enhancement. The numerical simulations are based on the full expressions derived in the previous section.
For all scenarios, we will use an input mirror of reflectivity $R_1 = 0.98$, a realistic value given current coating technology, accounting for both the finite transmission and the light losses in the $4f$ imaging optics left of the sample plane. For the output mirror, we shall assume $R_2 = 0.98 (1-T_2)$, either with variable transmission $T_2>0$ to control the output light in the CW and RD scenario, or with negligible $T_2 \ll 0.01$ to minimize roundtrip losses in the MP case.

\subsection{\label{sec:rayCW} Continuous wave cavity microscopy}

Formally, the continuous wave (CW) scenario corresponds to the stationary limit of a constant fixed-frequency input power $P_{\rm in}$. In practice, it is achieved in the limit of very long input pulses, such that the constructive interference over all round trips can lead to an enhanced intra-cavity field. The CW description applies if in addition also fringe effects due to the initial buildup and the final decay of the cavity enhancement are small, \ie{} the mean pulse duration $\tau$ far exceeds the inverse linewidth of the $8f$ cavity system. The time-integrated detection signal transmitted by the cavity and the energy absorbed by the sample can then be expressed in terms of the input power times a detection window. 
A further division by $\hbar \omega$ results in the respective photon numbers and in the dimensionless SNR evaluated below.

We expect a high sensitivity to the phase shift and absorption of a weak optical sample if the cavity is of high finesse, \ie{} supports many roundtrips. 
The empty imaging cavity has its resonances where $16kf$ is a multiple of the wavelength $\lambda$, and it supports the mean number of roundtrips $\la \ell \ra = \sqrt{R_1 R_2}/(1-\sqrt{R_1 R_2})$. 
A non-reflective sample holder shifts the resonances to $k(8f-d_g) = K\pi$. 
When the input field is tuned close to or on resonance, any sample-induced phase shift or extinction will result in a sharp change of the transmitted output signal, according to the Lorentz function. This is nicely illustrated in the WS approximation, where we can expand the stationary output field \eqref{eq:Eout_weak_w} around resonance. Omitting the glass plate and considering the limit of highly reflective mirrors at odd order $K$, we obtain to lowest order
\begin{eqnarray}
\frac{E_{\rm out}}{E_{\rm in}} &\approx& 
- \frac{t_1 t_2}{1-r_1 r_2} \lr 1 + \frac{4i\chi}{1-r_1 r_2} \rr, \label{eq:cav_resShift_Taylor}
\end{eqnarray}
This illustrates how the sample response is enhanced by the cavity finesse, \ie{} the number of supported roundtrips $\la \ell \ra \approx 1/(1-\sqrt{R_1 R_2}) \gg 1$. Given the detected light of an empty pixel, $Q_{\rm ref} \approx T_1 T_2 \la \ell \ra^2 Q_{\rm in} $,
the bright-field and Zernike signals reduce to $Q_{\rm BF} \approx 8\la \ell \ra \chii Q_{\rm ref}$ and $Q_{\rm Znk\pm} \approx \pm 8\la \ell \ra \chir Q_{\rm ref}$. The dark-field signal becomes $Q_{\rm DF} \approx 16 \la \ell \ra^2 |\chi|^2 Q_{\rm ref}$, and the absorbed energy simplifies to $Q_{\rm abs} \approx 8 \chii Q_{\rm ref}/T_2$. 
For weak samples, we thus find a common enhancement of the SNR in proportion to $\la \ell \ra \sqrt{T_2} $ at constant damage. Notice that this scaling factor reduces to $\sqrt{2 \la \ell \ra}$ if the input mirror is set to be almost perfect and other intra-cavity losses are neglected.

\begin{figure*}{
    \includegraphics[width=\textwidth]{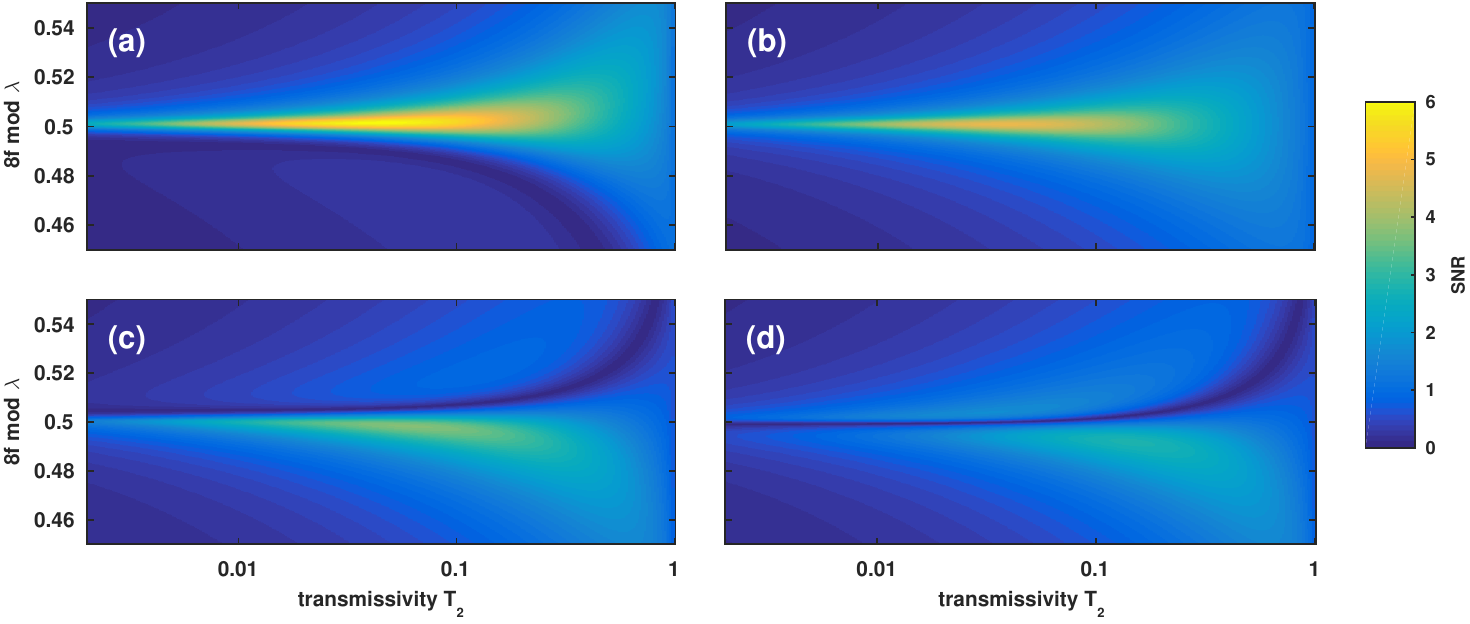}
    \caption{SNR for detecting a graphene monolayer with CW light as a function of the effective cavity length and end mirror transmissivity. We compare the BF readout (a), DF imaging (b), and the phase contrast schemes Znk+ (c) and Znk- (d). For all plots the damage was kept constant at the level obtained in a single interaction with a pulse of energy $1000\hbar\omega$, which corresponds to about $26$ absorbed photons.} \label{fig:CW1}}
\end{figure*}

\figref{fig:CW1} shows the SNR obtained for the detection of a graphene monolayer as a function of the effective cavity length (or detuning) and of $T_2$. (a-d) show the results obtained in a bright-field (BF), dark-field (DF), negative (Znk-) and positive (Znk+) phase contrast detection scheme, respectively. 
For all these plots the damage was kept constant at about $26$ absorbed photons, which is the damage that a short pulse of energy $T_1 Q_{\rm in}=1000\hbar\omega$ does in a single pass through the sample. The best SNR is found on cavity resonance, which gets more pronounced for lower $T_2$, corresponding to a cavity of higher quality.

\begin{figure}{\includegraphics[width=\columnwidth]{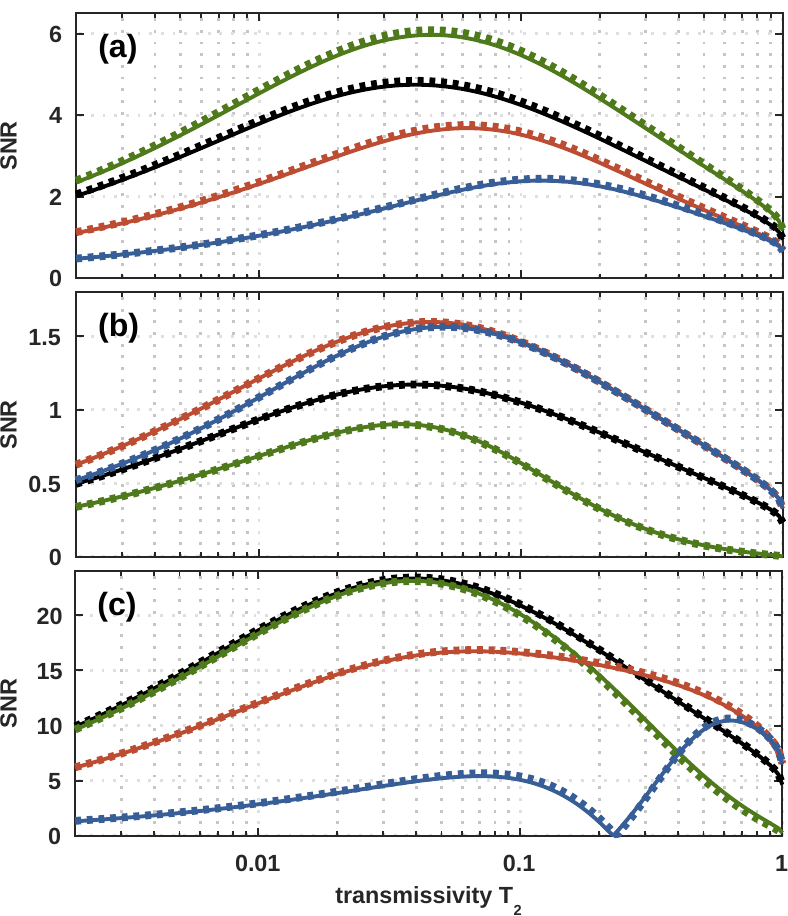}
    \caption{(a) SNR at constant damage as a function of output mirror transmissivity $T_2$ for CW detection of a graphene monolayer using BF imaging (green), DF (black), Znk+ (red), and Znk- (blue). For each scheme, we chose the cavity length with the highest maximum SNR (horizontal cut in \figref{fig:CW1} at $8f \mod \lambda \approx 0.5$), at the same damage level as before. Panels (b) and (c) show the SNR for detecting 1 and 20 monolayers of BN, respectively.} \label{fig:CW2}}
\end{figure}

In \figref{fig:CW2} (a) we provide a horizontal cut through the previous diagrams at a fixed detuning and plot the SNR as a function of $T_2$. For each of the four detection schemes, we chose the cavity length that supports the maximum SNR in \figref{fig:CW1}. The graphene monolayer yields the highest SNR in a BF detection scheme (green), followed by DF (black), Znk- (blue) and Znk+ (red). The dotted lines represent the WS approximation based on the output field expression \eqref{eq:Eout_weak_w}, which matches remarkably well even at low $T_2$ when the cavity supports many roundtrips. 

For comparison, we list the SNR values for single-pass detection at the same damage level in Tab.~\ref{tab:1}. A cavity with the mentioned specifications enhances the detection SNR by up to a factor of ten as compared to the optimal single pass microscopy technique. Even for $T_2=1$ the cavity simulations differ from these results due to light reflected from the specimen. 

\begin{table}
\begin{ruledtabular}
\begin{tabular}
{l*{4}{c}r}
             & C monolayer & BN monolayer & 20 BN monolayers \\
\hline
BF & $5.86\cdot10^{-1}$ & $3.07\cdot10^{-4}$ & $1.22\cdot10^{-1}$   \\
DF     & $4.75\cdot10^{-1}$ & $1.17\cdot10^{-1}$ & $2.34$   \\
Znk+   & $3.3\cdot10^{-1}$ & $1.66\cdot10^{-1}$ & $3.29$   \\
Znk-   & $3.22\cdot10^{-1}$ & $1.66\cdot10^{-1}$ & $3.29$   \\
\end{tabular}
\caption{SNR for the detection of 1 graphene monolayer, 1 BN monolayer, and 20 BN monolayers in conventional single-pass microscopy, where an input pulse of energy $1000\hbar\omega$ interacts with the specimen only once.}\label{tab:1}
\end{ruledtabular}
\end{table}

The results for a monolayer of BN are shown in \figref{fig:CW2} (b), keeping the damage level again fixed at the number of absorbed photons in a single interaction with a pulse energy $T_1 Q_{\rm in}=1000\hbar\omega$. We use the same reference value for all detection schemes and in all the following. 
Since BN has a negligible imaginary component of the refractive index, the number of absorbed photons will now be less than one. Hence for every $T_2$ in \figref{fig:CW2}, the SNR is evaluated at about the same mean number of light-sample interactions, albeit at a varying damage level in each sample. 
Given the real index of refraction, the best SNR in the BN case (b) is obtained in phase contrast readout schemes. The BF detection scheme gives the worst SNR. For a range of values of $T_2$ around 0.04 also BF gives a considerable SNR mainly because the interaction with the stationary cavity field translates phase to amplitude contrast. For 20 monolayers of BN in \figref{fig:CW2} (c) the detection SNR is generally higher. However, the accumulated phase shifts can now be significant, 
which is why BF and DF detection can out-compete phase contrast detection schemes, and also why the SNR in positive phase contrast readout goes to zero for $T_2 \sim 0.23$. Yet the WS approximation (dotted) still captures this behavior well.

\begin{figure}{\includegraphics[width=\columnwidth]{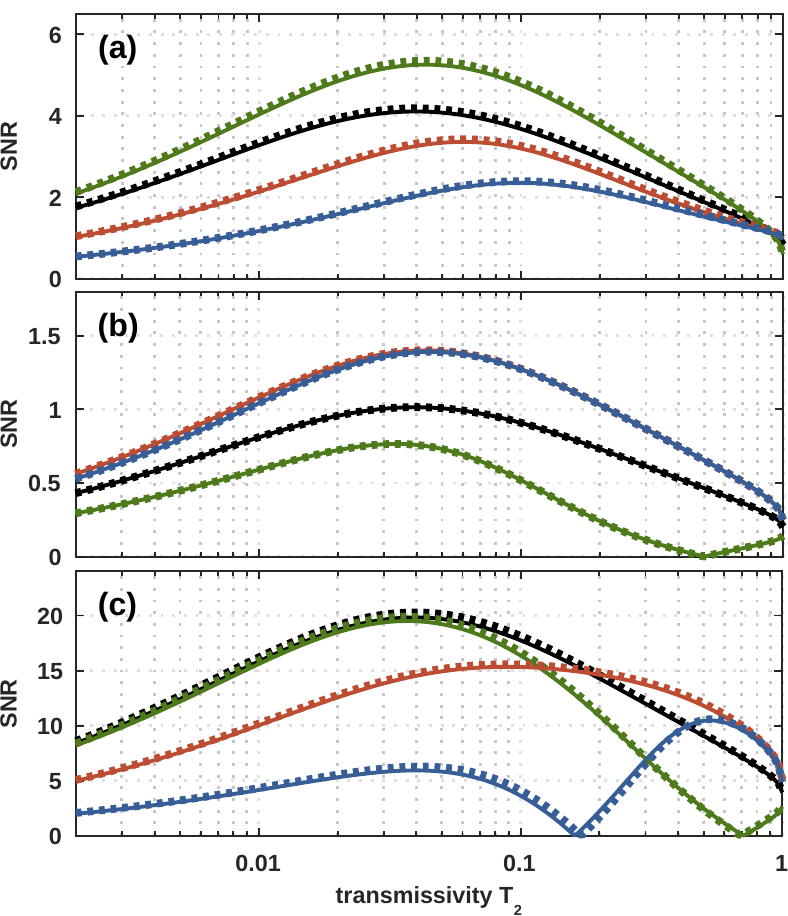}
    \caption{SNR at constant damage as in \figref{fig:CW2}, but with a sample holder of thickness $n_g k d_g = \pi$ at $n_g = 1.5$. Once again, we chose the optimal cavity length for BF (green), DF (black), Znk+ (red), and Znk- (blue). It is now at $(8f-d_g) \mod \lambda \approx 0$. We compare the results for a graphene monolayer (a), a BN monolayer (b), and 20 BN monolayers (c). In (b), both Znk schemes give the same curve.} \label{fig:CW3}}
\end{figure}

The above calculations were done for free sample layers without glass carrier plates, $d_g=0$. 
For non-reflective glass slabs, \ie{} multiples of half-wavelengths in optical thickness, the cavity resonances are shifted, but the achievable SNR are similar. For direct comparison with \figref{fig:CW2}, we show the results for a specimen carrier with $n_g = 1.5$ and $n_g k d_g = \pi$ in \figref{fig:CW3}. We remark that for $d_g=0$ the best SNR values are obtained close to an odd cavity resonance, $[ 8f \mod \lambda] \approx 0.5$, whereas now they are slightly lower and situated at $[(8f-d_g) \mod \lambda] \approx 0$. The glass plate induces an effective phase shift $2kd_g$ between the left- and right-running components that not only shifts the cavity resonance, but also modulates the reflections at the sample layer, as seen explicitly in \eqref{eq:Eout_weak_t}. In the pulsed imaging schemes discussed below, this will mainly affect the sample-reflected pulses outcoupled after an even number of sample interactions, see \eqref{eq:BF_pulse_even}.

A qualitatively different sample image would be observed if light were reflected by the carrier plate itself, \ie{} for $n_g k d_g \neq j\pi$. The weak response of the specimen would then be interlaced with the signature of the semi-transparent carrier, which typically results in a lower SNR for weak samples. We do not discuss this regime here.

\subsection{\label{sec:rayRing}Cavity ring down microscopy}

After the stationary scenario, where the light is allowed to interfere constructively in the imaging cavity, we now discuss the contrary regime of short, non-overlapping probe pulses. The straightforward way to enhance the sample signal by multiple sample interactions is a ring-down (RD) scheme \cite{OKeefe1988,2009CavitySpectroscopy}. A single input pulse of energy $Q_{\rm in}$ and temporal width $\tau < 8f/c$ is sent through the imaging cavity, accumulating losses and phase shifts as it bounces between the mirrors. The resulting output field is an attenuating train of pulses spaced by $\delta t=8f/c$, which can be either deposited as a cumulative signal in an integral detector or recorded individually in a time-resolved manner. 

We will first discuss the performance of the time-integrated RD scheme. The experimentally more demanding time-resolved detection of individual pulses provides more options for signal analysis, and it can be seen as a serial implementation of the multi-pass (MP) scheme discussed in Sect.~\ref{sec:raymulti-pass}. Each subsequent pulse corresponds to an increasing number of sample passes, but only a small fraction of it is then transmitted through the cavity end mirror, $T_2 \ll 1$. 

In order to assess the performance of RD imaging with respect to a conventional single-pass image, we vary the effective number of passes by tuning the transmission $T_2$ of the exit mirror and again adjust the input pulse energy accordingly to keep the total sample damage accumulated over all passes fixed, $m\to \infty$ in \eqref{eq:damage_m}. 
The BF, DF, and Zernike detection signals are given by the sums of the individual pulses from $m=0$ to $\infty$ in \eqref{eq:QbrightField}, \eqref{eq:IdarkField}, and \eqref{eq:IZernike}, respectively. For the BF and Zernike signal, we also subtract an empty reference pixel; its output signal $Q_{\rm ref} = Q_{\rm in} T_1 T_2/(1-R_1 R_2)$ then contributes to the noise. 

\begin{figure}{
	\includegraphics[width=\columnwidth]{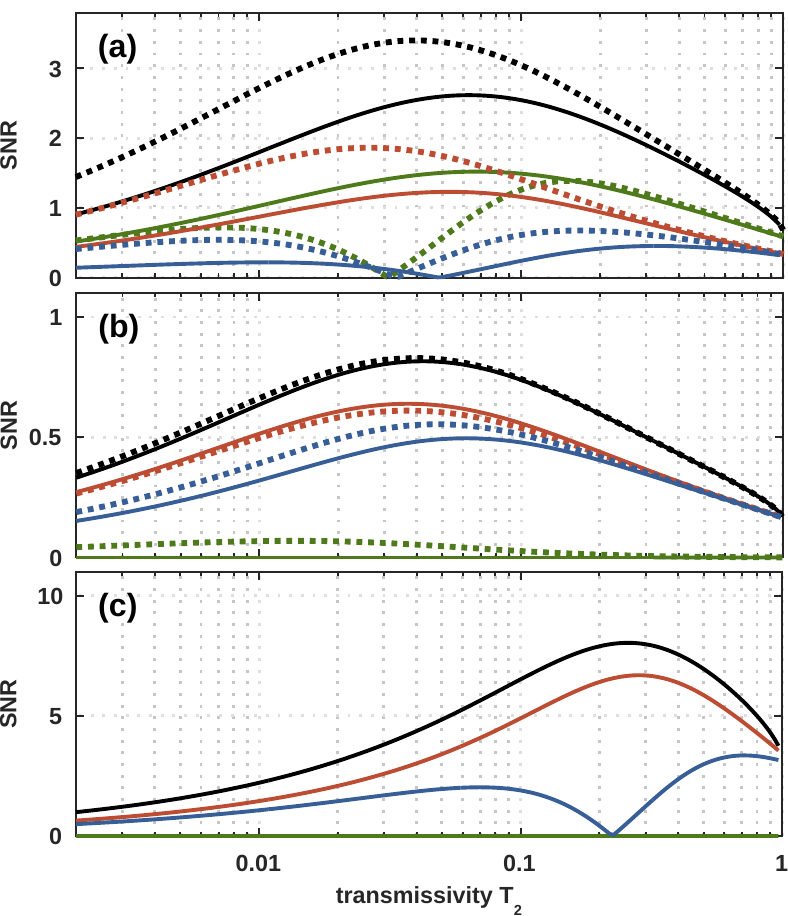} 
    \caption{SNR at constant damage as a function of output mirror transmissivity $T_2$ for RD detection of (a) a monolayer of graphene, (b) a monolayer of BN, and (c) 20 monolayers of BN. As in the CW case in \figref{fig:CW2} we compare BF imaging (green), DF (black), Znk+ (red), and Znk- (blue). The dotted lines show the WS approximation, which is omitted in (c) as it already diverges from the full result at $T_2 < 0.5$.} \label{fig:RD}}
\end{figure}

In the WS limit \eqref{eq:sampleWS}, we can compare the performance of the RD and the CW scheme by looking at the scaling with the empty-cavity roundtrip number $\la \ell \ra$ at high mirror reflectivities, $\la \ell \ra \approx 2/(1-R_1 R_2)$. The accumulated sample damage \eqref{eq:damage_mWS} can be expressed as $Q_{\rm abs}^{(\infty)} \approx 4 \chii Q_{\rm ref}/T_2$. The accumulated signals, on the other hand, reduce to 
$Q_{\rm BF} \approx 2 \la \ell \ra \chii Q_{\rm ref}$, $Q_{\rm Znk\pm} \approx \pm 2 \la \ell \ra \chir Q_{\rm ref}$, and $Q_{\rm DF} \approx 4 \la\ell\ra^2 |\chi|^2 Q_{\rm ref}$, as follows after summing Eqs.~\eqref{eq:BF_pulse_odd}, \eqref{eq:IdarkFieldWS}, and \eqref{eq:IZernikeWS} over all numbers $m$ of sample interactions. 

Notice that all the signals are four times smaller than their CW counterparts, while $Q_{\rm abs}$ is only two times smaller. Hence, at equal damage in the WS and high-reflectivity limit, the RD signals are by a factor of two worse than the CW signals and the SNR drops by $2\sqrt{2}$. This interferometric advantage is due to the coherent amplification of the intra-cavity field amplitudes that are transmitted and reflected by the sample in the CW case. Even in the limit of $T_1 \to 1$, we find that the linear sample response differs by a factor of two between the CW output field \eqref{eq:Eout_weak_w} and the first output pulse in \eqref{eq:Eout_weak_t}. The CW advantage comes with the experimental difficulty of having to stabilize the cavity at its resonance.  Nevertheless, the SNR scaling with $\sqrt{\la\ell\ra}$ remains the same in both schemes.

The numerical results are plotted in \figref{fig:RD} for the same samples and damage levels as in \figref{fig:CW2}. While there are many similarities to the previous CW case there are also some striking differences: First, the achieved SNR is consistently smaller, as discussed before. Moreover, the WS approximation, based on linearized expressions for the absorbed and detected pulse energies, quickly ceases to be valid as $T_2$ decreases. 
For the BN samples in \figref{fig:RD} (b) and (c), which are characterized by a real index of refraction, we notice that the time-integrated BF signal vanishes. In this case, the sample acts like a lossless beam splitter that redistributes the incident light energy over a trail of multiply reflected and transmitted pulses. Hence the time-integrated output energy is conserved and equal to the reference signal. This BF signal cancellation could be avoided with time-tagged detection, \eg{} by using an avalanche photodiode array detector \cite{Richardson2009AImaging} flipping the sign of subsequent pulses in the post-processing stage. The same technique would also avert SNR cancellation in the negative phase contrast detection of 20 BN monolayers, which is reflected in the dip of the blue curve in \figref{fig:RD}(c). 

The results were again evaluated without sample holder, $d_g=0$. Using a non-reflective glass plate as in \figref{fig:CW3} for the CW case, the achievable SNR values in RD imaging would also exhibit a slightly different $T_2$-dependence and overall decrease. This is mainly due to the suppression of sample reflections, as seen explicitly in the WS limit. There, only the (weaker) even pulse orders \eqref{eq:BF_pulse_even} are affected by the sample holder, in proportion to $2\sqrt{1-T_2} \cos (2kd_g) - T_2$. One thus obtains slightly better SNR values if $kd_g$ is zero or a multiple of $\pi$. 

\subsection{\label{sec:raymulti-pass}Multi-pass microscopy}

In multi-pass (MP) imaging the goal is to limit and control the number of sample interactions of a short pulse entering a high-finesse imaging cavity. 
This can be achieved, for instance, by placing a fast outcoupling mechanism behind the sample plane that is locked to the input pulse timing and triggers after a delay corresponding to a selected number $\ell$ of full roundtrips. 
This allows one to choose between odd numbers of sample interactions, $m=2\ell+1$. 
Ideally, the outcoupling occurs at unit efficiency, while the imaging cavity should be of high finesse to minimize any sample-independent roundtrip losses. In practice, one can implement the outcoupling by means of a Pockels cell and a polarizing beam splitter. A Pockels cell is routinely incorporated in optical cavities and its losses can be neglected compared to realistic losses at lens interfaces. 
For convenience, we shall assume $R_1=R_2=R$, using $R=0.98$ for the numerical examples. The empty reference pixel yields the signal $Q_{\rm ref} = T R^{2\ell} Q_{\rm in}$.
(Note that one could also outcouple after even numbers of sample interactions, which captures the fraction of light reflected at the sample. The outcoupled light would not contain the bright background contribution $Q_{\rm ref}$, but rather resemble the DF image after full roundtrips.)

Once again, we can estimate the explicit scaling of the SNR with the selected number of roundtrips in the WS limit. Here, the damage reduces to $Q_{\rm abs}^{(2\ell+1)} \approx 2 T Q_{\rm in} \chii (1-R^{2\ell+1})/(1-R)$. The BF and Z signals follow from \eqref{eq:BF_pulse_odd} and \eqref{eq:IZernikeWS} after removing $T_2$ and subtracting the empty pixel, $Q_{\rm BF} \approx 2 \chii  Q_{\rm ref} (2\ell+1) $ and $Q_{\rm Znk\pm} \approx \pm 2 \chir Q_{\rm ref} (2\ell+1) $, whereas $Q_{\rm DF} \approx |\chi|^2 Q_{\rm ref} (2\ell+1)^2$. Dividing by the respective shot noise amplitudes leaves us with an SNR that grows with the number of passes like $\sqrt{(1-R)/(1-R^{2\ell+1})} R^\ell (2\ell+1)$. For a moderate number of roundtrips, we can expand to lowest order in $\varepsilon = 1-R$ and find a square-root enhancement, $\SNR_{\rm bf,df,Z} \propto \sqrt{2\ell +1}$. After many more roundtrips, the SNR decreases again exponentially with $R^\ell \sim e^{-\ell \varepsilon}$. So we expect a sweet spot of maximum SNR at roughly $\ell \sim 1/\varepsilon$, provided the extinction at the weak sample is lower than the cavity loss and the accumulated phase shift $(2\ell+1)\chir \ll \pi$.

\begin{figure}{
	\includegraphics[width=\columnwidth]{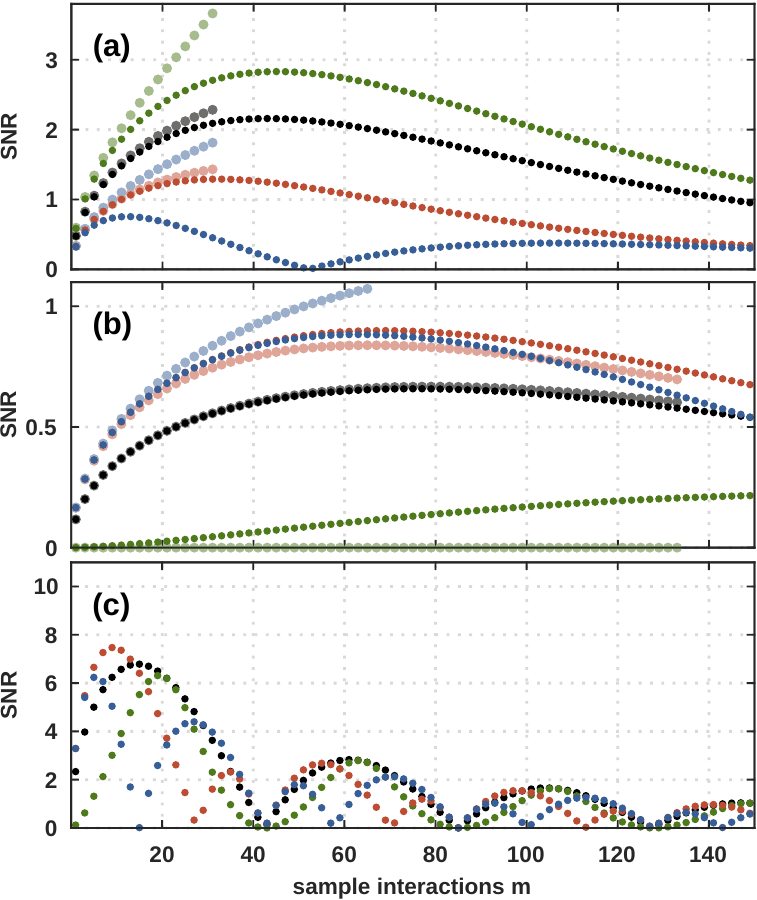}
	\caption{SNR at constant damage for MP detection with a probe pulse that passes multiple times through (a) a monolayer of graphene, (b) a monolayer of BN, and (c) 20 monolayers of BN. We assume perfect outcoupling efficiency after each odd multiple $m$. The green, black, red, and blue dots correspond to BF, DF, Znk+, and Znk- imaging, respectively. In (a) and (b), we plot the WS approximation (bright circles) up to $2m|\chi| = 1$, where it is certainly no more valid.} \label{fig:MP}}
\end{figure}

The achievable SNR as a function of $m$ is plotted in \figref{fig:MP} (small dots). The damage levels were again chosen as in the CW and RD case. At that illumination intensity the single-pass SNR for detecting a monolayer of graphene or BN is below unity, irrespective of whether the sample is investigated a BF (green), DF (black), Znk+ (red) or Znk-(blue) microscope, see leftmost data points in \figref{fig:MP} (a) and (b). Multiple passes initially increase the SNR, until losses eventually outweigh the gain in sensitivity offered by each additional pass. 
Due to the negligible loss in BN, its optimum sensitivity is reached after a higher number of roundtrips than for graphene. 
Given that all the light can be outcoupled after the optimal number of interactions, it might be surprising that the achievable SNR are very close to the ones observed in \figref{fig:RD} for the RD case. This is because the light that underwent an odd number of reflections from (\ie{} even number $m$ of interactions with) the sample forms a counter-propagating pulse that is not outcoupled to the detector. It thus neither cancels the BF signals, as observed in a RD scheme for the BN samples, nor does it contribute to the DF or phase-contrast signals. 
The performance of a MP scheme can potentially be improved by adjusting the timing window of the detection to also include the counter-propagating pulse. 
 \figref{fig:MP} (c) shows the signal for 20 monolayers of BN. The obtained SNR for this thicker sample is higher and shows oscillations as the phase shifts accumulate. 
The WS approximation (bright circles) is valid for a several tens of passes in (a) and (b); it does not hold for the 20 monolayers thick sample and is therefore omitted in (c). We note that samples of sub-diffraction limited dimensions would represent even weaker samples, for which the WS limit might be appropriate also at high interaction numbers.

\begin{figure}{
	\includegraphics[width=\columnwidth]{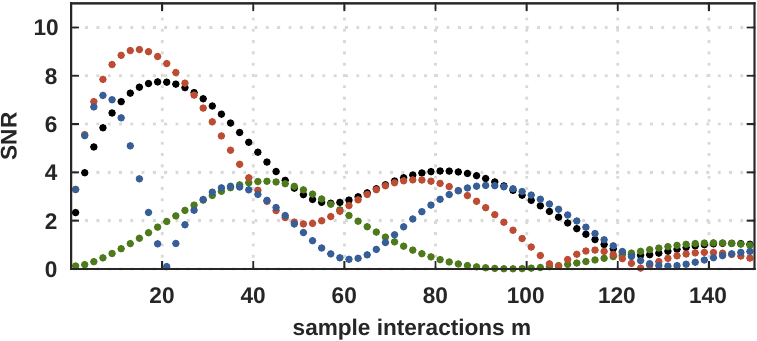}
	\caption{SNR at constant damage for MP detection of 20 monolayers of BN on a glass plate of optical thickness $n_g k d_g = \pi$ at $n_g=1.5$. The green, black, red, and blue dots represent BF, DF, Znk+, and Znk- imaging, respectively. Compare to \figref{fig:MP} (c) without glass plate.} \label{fig:MP2}}
\end{figure}

In the case of a finite carrier plate, $n_g k d_g = \pi$ as studied in Sect.~\ref{sec:rayCW}, we expect (and numerical simulations confirm) no significant change in the SNR associated to odd interaction orders $m$ that are plotted for the single monolayers in \figref{fig:MP} (a) and (b). The behavior does change for 20 BN monolayers (c) where the WS approximation is no more valid. We plot the results with sample carrier in \figref{fig:MP2}. 

Note that the picture changes completely if we consider reflective sample holders, $n_g k d_g \neq j\pi$. Light that has undergone multiple reflections and transmissions at the sample would then be redistributed over several pulses by the carrier plate, which acts as a semi-transparent mirror. No clear distinction between even and odd $m$ could be made in terms of the sample response any longer.

\section{Conclusions \label{sec:Conclusions}}

It is a well established result in the quantum description of phase estimation~\cite{Luis2002a,Giovannetti2006a,Higgins2007a,Braun2017QuantumEntanglement} that multiple interactions between a probe particle and a specimen can lead to an enhanced measurement precision beyond the shot-noise limit. The same enhancement can be observed in optical wide-field microscopy, using self-imaging cavities to amplify the phase and amplitude contrast of optically thin, weak samples. Here we have assessed in detail how the signal-to-noise ratio increases with the number of light-sample interactions. Our results not only apply to the recently demonstrated pulsed multipass schemes with a controlled number of light-sample interactions \cite{Juffmann2016a,Juffmann2016Multi-passMicroscopyb}, but also to ring-down and continuous-wave schemes where the enhanced interaction strength is set by the finesse of the self-imaging cavity. 

Indeed, we find that in the limit of weak samples and loss-less cavities, the sensitivity of appropriate (optimal) detection schemes always grows with the square root of the mean number of light-sample interactions---in agreement with results from quantum measurement theory \cite{Giovannetti2006a}.
When the detection scheme is not ideal with respect to the properties of the specimen, the sensitivity enhancement can even surpass the square-root scaling, as we have shown for bright-field images of non-absorbing phase shifters, for instance. This might be a desireable feature if the detection technique can not be altered due to other experimental constraints, or if the sample properties are completely unknown in the first place. 

In the presented case studies, the detection of a few atomic layers of graphene or boron nitride, all cavity-enhanced microscopy techniques could clearly outperform a conventional single-pass image in terms of signal-to-noise at a fixed number of photon-specimen interactions. We obtained the best results with the continuous-wave approach, which profits from the interference between co- and counter-propagating components of the intra-cavity field that does not occur in a pulsed multipass or ring-down scheme. 
The resonance behavior also leads to a conversion from phase contrast to amplitude contrast, such that a pure phase sample can be detected using a bright-field detection scheme. 
However, the sensitivity enhancement requires the imaging cavity to be fine-tuned to its sample-dependent resonance and is thus more challenging to realize in an experiment. 

In the pulsed multipass and ring-down schemes, co- and counter-propagating light pulses will reach the detector at different times. Even though no interference takes place between those pulses, the phase or amplitude signal of the sample can still be amplified, de-amplified, or cancelled, depending on the chosen imaging technique. For further improving the signal-to-noise ratio, one can employ time-gated detection (\eg{} using an avalanche photodiode array detector \cite{Richardson2009AImaging}), or consider ring resonators that outcouple co- and counter-propagating pulses in different directions and onto separate detectors. 

While the analysis was done for scalar fields, it was shown that polarization-sensitive measurement techniques are also enhanced by multi-passing \cite{Juffmann2016a}, with potential applications such as ellipsometry. 
Further numerical studies may give more insights to the diffraction-limited detection of sub-wavelength samples such as nano-structured materials or biological specimen. First results were only obtained for continuous-wave imaging via the transverse modes of a resonant, degenerate cavity \cite{Gigan2005}. It remains to be investigated how the less stringent requirements of a pulsed detection scheme would affect the signal, and possibly the resolution and depth of field obtained when imaging sub-wavelength structures.

\begin{acknowledgments}
This research is funded by the Gordon and Betty Moore Foundation. B.B.K.~is supported by a Stanford Graduate Fellowship. T.J.~is supported by the Human Frontier Science Program. S.N.~is supported by the National Research Foundation, Prime Minister's Office, Singapore, and the Ministry of Education, Singapore under the Research Centres of Excellence programme.
\end{acknowledgments}

%Bibliography 


\begin{thebibliography}{34}%
\makeatletter
\providecommand \@ifxundefined [1]{%
 \@ifx{#1\undefined}
}%
\providecommand \@ifnum [1]{%
 \ifnum #1\expandafter \@firstoftwo
 \else \expandafter \@secondoftwo
 \fi
}%
\providecommand \@ifx [1]{%
 \ifx #1\expandafter \@firstoftwo
 \else \expandafter \@secondoftwo
 \fi
}%
\providecommand \natexlab [1]{#1}%
\providecommand \enquote  [1]{``#1''}%
\providecommand \bibnamefont  [1]{#1}%
\providecommand \bibfnamefont [1]{#1}%
\providecommand \citenamefont [1]{#1}%
\providecommand \href@noop [0]{\@secondoftwo}%
\providecommand \href [0]{\begingroup \@sanitize@url \@href}%
\providecommand \@href[1]{\@@startlink{#1}\@@href}%
\providecommand \@@href[1]{\endgroup#1\@@endlink}%
\providecommand \@sanitize@url [0]{\catcode `\\12\catcode `\$12\catcode
  `\&12\catcode `\#12\catcode `\^12\catcode `\_12\catcode `\%12\relax}%
\providecommand \@@startlink[1]{}%
\providecommand \@@endlink[0]{}%
\providecommand \url  [0]{\begingroup\@sanitize@url \@url }%
\providecommand \@url [1]{\endgroup\@href {#1}{\urlprefix }}%
\providecommand \urlprefix  [0]{URL }%
\providecommand \Eprint [0]{\href }%
\providecommand \doibase [0]{http://dx.doi.org/}%
\providecommand \selectlanguage [0]{\@gobble}%
\providecommand \bibinfo  [0]{\@secondoftwo}%
\providecommand \bibfield  [0]{\@secondoftwo}%
\providecommand \translation [1]{[#1]}%
\providecommand \BibitemOpen [0]{}%
\providecommand \bibitemStop [0]{}%
\providecommand \bibitemNoStop [0]{.\EOS\space}%
\providecommand \EOS [0]{\spacefactor3000\relax}%
\providecommand \BibitemShut  [1]{\csname bibitem#1\endcsname}%
\let\auto@bib@innerbib\@empty
%</preamble>
\bibitem [{\citenamefont {Mader}\ \emph {et~al.}(2015)\citenamefont {Mader},
  \citenamefont {Reichel}, \citenamefont {Hansch},\ and\ \citenamefont
  {Hunger}}]{Mader2015a}%
  \BibitemOpen
  \bibfield  {author} {\bibinfo {author} {\bibfnamefont {M.}~\bibnamefont
  {Mader}}, \bibinfo {author} {\bibfnamefont {J.}~\bibnamefont {Reichel}},
  \bibinfo {author} {\bibfnamefont {T.~W.}\ \bibnamefont {Hansch}}, \ and\
  \bibinfo {author} {\bibfnamefont {D.}~\bibnamefont {Hunger}},\ }\href
  {http://dx.doi.org/10.1038/ncomms8249} {\bibfield  {journal} {\bibinfo
  {journal} {Nat.~Commun.}\ }\textbf {\bibinfo {volume} {6}} (\bibinfo {year}
  {2015})}\BibitemShut {NoStop}%
\bibitem [{\citenamefont {H{\"{u}}mmer}\ \emph {et~al.}(2015)\citenamefont
  {H{\"{u}}mmer}, \citenamefont {Noe}, \citenamefont {Hofmann}, \citenamefont
  {H{\"{a}}nsch}, \citenamefont {H{\"{o}}gele},\ and\ \citenamefont
  {Hunger}}]{Hummer2015Cavity-enhancedNanotubes}%
  \BibitemOpen
  \bibfield  {author} {\bibinfo {author} {\bibfnamefont {T.}~\bibnamefont
  {H{\"{u}}mmer}}, \bibinfo {author} {\bibfnamefont {J.}~\bibnamefont {Noe}},
  \bibinfo {author} {\bibfnamefont {M.~S.}\ \bibnamefont {Hofmann}}, \bibinfo
  {author} {\bibfnamefont {T.~W.}\ \bibnamefont {H{\"{a}}nsch}}, \bibinfo
  {author} {\bibfnamefont {A.}~\bibnamefont {H{\"{o}}gele}}, \ and\ \bibinfo
  {author} {\bibfnamefont {D.}~\bibnamefont {Hunger}},\ }\href {\doibase
  10.1038/ncomms12155} {\  (\bibinfo {year} {2015}),\
  10.1038/ncomms12155}\BibitemShut {NoStop}%
\bibitem [{\citenamefont {Tolansky}(1970)}]{tolansky1970multiple}%
  \BibitemOpen
  \bibfield  {author} {\bibinfo {author} {\bibfnamefont {S.}~\bibnamefont
  {Tolansky}},\ }\href {https://books.google.fr/books?id=hhkzAAAAMAAJ} {\emph
  {\bibinfo {title} {{Multiple-beam interference microscopy of metals}}}}\
  (\bibinfo  {publisher} {Academic Press},\ \bibinfo {year} {1970})\BibitemShut
  {NoStop}%
\bibitem [{\citenamefont {Juffmann}\ \emph
  {et~al.}(2016{\natexlab{a}})\citenamefont {Juffmann}, \citenamefont
  {Klopfer}, \citenamefont {Frankort}, \citenamefont {Haslinger},\ and\
  \citenamefont {Kasevich}}]{Juffmann2016a}%
  \BibitemOpen
  \bibfield  {author} {\bibinfo {author} {\bibfnamefont {T.}~\bibnamefont
  {Juffmann}}, \bibinfo {author} {\bibfnamefont {B.~B.}\ \bibnamefont
  {Klopfer}}, \bibinfo {author} {\bibfnamefont {T.~L.~I.}\ \bibnamefont
  {Frankort}}, \bibinfo {author} {\bibfnamefont {P.}~\bibnamefont {Haslinger}},
  \ and\ \bibinfo {author} {\bibfnamefont {M.~A.}\ \bibnamefont {Kasevich}},\
  }\href@noop {} {\bibfield  {journal} {\bibinfo  {journal} {submitted.}\ }
  (\bibinfo {year} {2016}{\natexlab{a}})}\BibitemShut {NoStop}%
\bibitem [{\citenamefont {Klopfer}\ \emph {et~al.}(2016)\citenamefont
  {Klopfer}, \citenamefont {Juffmann},\ and\ \citenamefont
  {Kasevich}}]{Klopfer2016IterativeLight}%
  \BibitemOpen
  \bibfield  {author} {\bibinfo {author} {\bibfnamefont {B.~B.}\ \bibnamefont
  {Klopfer}}, \bibinfo {author} {\bibfnamefont {T.}~\bibnamefont {Juffmann}}, \
  and\ \bibinfo {author} {\bibfnamefont {M.~A.}\ \bibnamefont {Kasevich}},\
  }\href {\doibase 10.1364/OL.41.005744} {\bibfield  {journal} {\bibinfo
  {journal} {Opt.~Lett.}\ }\textbf {\bibinfo {volume} {41}},\ \bibinfo
  {pages} {5744} (\bibinfo {year} {2016})}\BibitemShut {NoStop}%
\bibitem [{\citenamefont {Hall}(1974)}]{Hall1974Multiple-BeamInterferometry}%
  \BibitemOpen
  \bibfield  {author} {\bibinfo {author} {\bibfnamefont {A.~C.}\ \bibnamefont
  {Hall}},\ }in\ \href {\doibase 10.1007/978-1-4613-4490-2{\_}3} {\emph
  {\bibinfo {booktitle} {Characterization of Solid Surfaces}}}\ (\bibinfo
  {publisher} {Springer US},\ \bibinfo {address} {Boston, MA},\ \bibinfo {year}
  {1974})\ pp.\ \bibinfo {pages} {33--48}\BibitemShut {NoStop}%
\bibitem [{\citenamefont {Arnaud}(1969)}]{Arnaud1969a}%
  \BibitemOpen
  \bibfield  {author} {\bibinfo {author} {\bibfnamefont {J.~A.}\ \bibnamefont
  {Arnaud}},\ }\href {http://ao.osa.org/abstract.cfm?URI=ao-8-1-189} {\bibfield
   {journal} {\bibinfo  {journal} {Appl.~Opt.}\ }\textbf {\bibinfo {volume}
  {8}},\ \bibinfo {pages} {189} (\bibinfo {year} {1969})}\BibitemShut {NoStop}%
\bibitem [{\citenamefont {Gigan}\ \emph
  {et~al.}(2005{\natexlab{a}})\citenamefont {Gigan}, \citenamefont {Lopez},
  \citenamefont {Treps}, \citenamefont {Ma{\^{i}}tre},\ and\ \citenamefont
  {Fabre}}]{Gigan2005a}%
  \BibitemOpen
  \bibfield  {author} {\bibinfo {author} {\bibfnamefont {S.}~\bibnamefont
  {Gigan}}, \bibinfo {author} {\bibfnamefont {L.}~\bibnamefont {Lopez}},
  \bibinfo {author} {\bibfnamefont {N.}~\bibnamefont {Treps}}, \bibinfo
  {author} {\bibfnamefont {A.}~\bibnamefont {Ma{\^{i}}tre}}, \ and\ \bibinfo
  {author} {\bibfnamefont {C.}~\bibnamefont {Fabre}},\ }\href
  {http://link.aps.org/doi/10.1103/PhysRevA.72.023804} {\bibfield  {journal}
  {\bibinfo  {journal} {Phys.~Rev.~A}\ }\textbf {\bibinfo {volume} {72}},\
  \bibinfo {pages} {23804} (\bibinfo {year} {2005}{\natexlab{a}})}\BibitemShut
  {NoStop}%
\bibitem [{\citenamefont {Giovannetti}\ \emph {et~al.}(2011)\citenamefont
  {Giovannetti}, \citenamefont {Lloyd},\ and\ \citenamefont
  {Maccone}}]{Giovannetti2011a}%
  \BibitemOpen
  \bibfield  {author} {\bibinfo {author} {\bibfnamefont {V.}~\bibnamefont
  {Giovannetti}}, \bibinfo {author} {\bibfnamefont {S.}~\bibnamefont {Lloyd}},
  \ and\ \bibinfo {author} {\bibfnamefont {L.}~\bibnamefont {Maccone}},\ }\href
  {http://dx.doi.org/10.1038/nphoton.2011.35} {\bibfield  {journal} {\bibinfo
  {journal} {Nat.~Photon.}\ }\textbf {\bibinfo {volume} {5}},\ \bibinfo {pages}
  {222} (\bibinfo {year} {2011})}\BibitemShut {NoStop}%
\bibitem [{\citenamefont {Braun}\ \emph {et~al.}(2017)\citenamefont {Braun},
  \citenamefont {Adesso}, \citenamefont {Benatti}, \citenamefont {Floreanini},
  \citenamefont {Marzolino}, \citenamefont {Mitchell},\ and\ \citenamefont
  {Pirandola}}]{Braun2017QuantumEntanglement}%
  \BibitemOpen
  \bibfield  {author} {\bibinfo {author} {\bibfnamefont {D.}~\bibnamefont
  {Braun}}, \bibinfo {author} {\bibfnamefont {G.}~\bibnamefont {Adesso}},
  \bibinfo {author} {\bibfnamefont {F.}~\bibnamefont {Benatti}}, \bibinfo
  {author} {\bibfnamefont {R.}~\bibnamefont {Floreanini}}, \bibinfo {author}
  {\bibfnamefont {U.}~\bibnamefont {Marzolino}}, \bibinfo {author}
  {\bibfnamefont {M.~W.}\ \bibnamefont {Mitchell}}, \ and\ \bibinfo {author}
  {\bibfnamefont {S.}~\bibnamefont {Pirandola}},\ }\href@noop {} {\  (\bibinfo
  {year} {2017})}\BibitemShut {NoStop}%
\bibitem [{\citenamefont {Giovannetti}\ \emph {et~al.}(2006)\citenamefont
  {Giovannetti}, \citenamefont {Lloyd},\ and\ \citenamefont
  {Maccone}}]{Giovannetti2006a}%
  \BibitemOpen
  \bibfield  {author} {\bibinfo {author} {\bibfnamefont {V.}~\bibnamefont
  {Giovannetti}}, \bibinfo {author} {\bibfnamefont {S.}~\bibnamefont {Lloyd}},
  \ and\ \bibinfo {author} {\bibfnamefont {L.}~\bibnamefont {Maccone}},\ }\href
  {http://link.aps.org/doi/10.1103/PhysRevLett.96.010401} {\bibfield  {journal}
  {\bibinfo  {journal} {Phys.~Rev.~Lett.}\ }\textbf {\bibinfo {volume}
  {96}},\ \bibinfo {pages} {10401} (\bibinfo {year} {2006})}\BibitemShut
  {NoStop}%
\bibitem [{\citenamefont {Luis}(2002)}]{Luis2002a}%
  \BibitemOpen
  \bibfield  {author} {\bibinfo {author} {\bibfnamefont {A.}~\bibnamefont
  {Luis}},\ }\href {http://link.aps.org/doi/10.1103/PhysRevA.65.025802}
  {\bibfield  {journal} {\bibinfo  {journal} {Phys.~Rev.~A}\ }\textbf
  {\bibinfo {volume} {65}},\ \bibinfo {pages} {25802} (\bibinfo {year}
  {2002})}\BibitemShut {NoStop}%
\bibitem [{\citenamefont {Higgins}\ \emph {et~al.}(2007)\citenamefont
  {Higgins}, \citenamefont {Berry}, \citenamefont {Bartlett}, \citenamefont
  {Wiseman},\ and\ \citenamefont {Pryde}}]{Higgins2007a}%
  \BibitemOpen
  \bibfield  {author} {\bibinfo {author} {\bibfnamefont {B.~L.}\ \bibnamefont
  {Higgins}}, \bibinfo {author} {\bibfnamefont {D.~W.}\ \bibnamefont {Berry}},
  \bibinfo {author} {\bibfnamefont {S.~D.}\ \bibnamefont {Bartlett}}, \bibinfo
  {author} {\bibfnamefont {H.~M.}\ \bibnamefont {Wiseman}}, \ and\ \bibinfo
  {author} {\bibfnamefont {G.~J.}\ \bibnamefont {Pryde}},\ }\href
  {http://dx.doi.org/10.1038/nature06257} {\bibfield  {journal} {\bibinfo
  {journal} {Nature}\ }\textbf {\bibinfo {volume} {450}},\ \bibinfo {pages}
  {393} (\bibinfo {year} {2007})}\BibitemShut {NoStop}%
\bibitem [{\citenamefont {W{\"{a}}ldchen}\ \emph {et~al.}(2015)\citenamefont
  {W{\"{a}}ldchen} \emph {et~al.}}]{Waldchen2015short}%
  \BibitemOpen
  \bibfield  {author} {\bibinfo {author} {\bibfnamefont {S.}~\bibnamefont
  {W{\"{a}}ldchen}} \emph {et~al.},\ }\href {\doibase 10.1038/srep15348}
  {\bibfield  {journal} {\bibinfo  {journal} {Sci.~Rep.}\ }\textbf
  {\bibinfo {volume} {5}},\ \bibinfo {pages} {15348} (\bibinfo {year}
  {2015})}\BibitemShut {NoStop}%
\bibitem [{\citenamefont {Juffmann}\ \emph
  {et~al.}(2016{\natexlab{b}})\citenamefont {Juffmann}, \citenamefont
  {Koppell}, \citenamefont {Klopfer}, \citenamefont {Ophus}, \citenamefont
  {Glaeser},\ and\ \citenamefont
  {Kasevich}}]{Juffmann2016Multi-passMicroscopyb}%
  \BibitemOpen
  \bibfield  {author} {\bibinfo {author} {\bibfnamefont {T.}~\bibnamefont
  {Juffmann}}, \bibinfo {author} {\bibfnamefont {S.~A.}\ \bibnamefont
  {Koppell}}, \bibinfo {author} {\bibfnamefont {B.~B.}\ \bibnamefont
  {Klopfer}}, \bibinfo {author} {\bibfnamefont {C.}~\bibnamefont {Ophus}},
  \bibinfo {author} {\bibfnamefont {R.}~\bibnamefont {Glaeser}}, \ and\
  \bibinfo {author} {\bibfnamefont {M.~A.}\ \bibnamefont {Kasevich}},\ }\href
  {http://arxiv.org/abs/1612.04931} {\  (\bibinfo {year}
  {2016}{\natexlab{b}})}\BibitemShut {NoStop}%
\bibitem [{\citenamefont {Glaeser}(2016)}]{Glaeser2016}%
  \BibitemOpen
  \bibfield  {author} {\bibinfo {author} {\bibfnamefont {R.~M.}\ \bibnamefont
  {Glaeser}},\ }\href {http://dx.doi.org/10.1038/nmeth.3695} {\bibfield
  {journal} {\bibinfo  {journal} {Nat.~Meth.}\ }\textbf {\bibinfo {volume}
  {13}},\ \bibinfo {pages} {28} (\bibinfo {year} {2016})}\BibitemShut {NoStop}%
\bibitem [{\citenamefont {Zernike}(1942)}]{Zernike1942}%
  \BibitemOpen
  \bibfield  {author} {\bibinfo {author} {\bibfnamefont {F.}~\bibnamefont
  {Zernike}},\ }\href {\doibase 10.1016/S0031-8914(42)80035-X} {\bibfield
  {journal} {\bibinfo  {journal} {Physica}\ }\textbf {\bibinfo {volume} {9}},\
  \bibinfo {pages} {686} (\bibinfo {year} {1942})}\BibitemShut {NoStop}%
\bibitem [{\citenamefont {Mertz}(2009)}]{mertz2009introduction}%
  \BibitemOpen
  \bibfield  {author} {\bibinfo {author} {\bibfnamefont {J.}~\bibnamefont
  {Mertz}},\ }\href {https://books.google.fr/books?id=djxnPgAACAAJ} {\emph
  {\bibinfo {title} {{Introduction to Optical Microscopy}}}}\ (\bibinfo
  {publisher} {W. H. Freeman},\ \bibinfo {year} {2009})\BibitemShut {NoStop}%
\bibitem [{\citenamefont {Orfanidis}(2002)}]{Orfanidis2002}%
  \BibitemOpen
  \bibfield  {author} {\bibinfo {author} {\bibfnamefont {S.~J.}\ \bibnamefont
  {Orfanidis}},\ }\href@noop {} {\emph {\bibinfo {title} {{Electromagnetic
  waves and antennas}}}}\ (\bibinfo  {publisher} {Rutgers University New
  Brunswick, NJ},\ \bibinfo {year} {2002})\BibitemShut {NoStop}%
\bibitem [{\citenamefont {Siegman}(1986)}]{Siegman1986Lasers}%
  \BibitemOpen
  \bibfield  {author} {\bibinfo {author} {\bibfnamefont {A.~E.}\ \bibnamefont
  {Siegman}},\ }\href@noop {} {\emph {\bibinfo {title} {{Lasers}}}}\ (\bibinfo
  {publisher} {University Science Books},\ \bibinfo {year} {1986})\ p.\
  \bibinfo {pages} {1283}\BibitemShut {NoStop}%
\bibitem [{\citenamefont {Saleh}\ and\ \citenamefont
  {Teich}(1991)}]{Saleh1991}%
  \BibitemOpen
  \bibfield  {author} {\bibinfo {author} {\bibfnamefont {B.~E.~A.}\
  \bibnamefont {Saleh}}\ and\ \bibinfo {author} {\bibfnamefont {M.~C.}\
  \bibnamefont {Teich}},\ }\href@noop {} {\emph {\bibinfo {title}
  {{Fundamentals of photonics}}}}\ (\bibinfo  {publisher} {Wiley},\ \bibinfo
  {address} {New York},\ \bibinfo {year} {1991})\BibitemShut {NoStop}%
\bibitem [{\citenamefont {Giovannetti}\ \emph {et~al.}(2004)\citenamefont
  {Giovannetti}, \citenamefont {Lloyd},\ and\ \citenamefont
  {Maccone}}]{Giovannetti2004a}%
  \BibitemOpen
  \bibfield  {author} {\bibinfo {author} {\bibfnamefont {V.}~\bibnamefont
  {Giovannetti}}, \bibinfo {author} {\bibfnamefont {S.}~\bibnamefont {Lloyd}},
  \ and\ \bibinfo {author} {\bibfnamefont {L.}~\bibnamefont {Maccone}},\ }\href
  {http://www.sciencemag.org/content/306/5700/1330.abstract} {\bibfield
  {journal} {\bibinfo  {journal} {Science}\ }\textbf {\bibinfo {volume}
  {306}},\ \bibinfo {pages} {1330} (\bibinfo {year} {2004})}\BibitemShut
  {NoStop}%
\bibitem [{\citenamefont {Cheon}\ \emph {et~al.}(2014)\citenamefont {Cheon},
  \citenamefont {Kihm}, \citenamefont {Kim}, \citenamefont {Lim}, \citenamefont
  {Park},\ and\ \citenamefont {Lee}}]{Cheon2014HowConstraints}%
  \BibitemOpen
  \bibfield  {author} {\bibinfo {author} {\bibfnamefont {S.}~\bibnamefont
  {Cheon}}, \bibinfo {author} {\bibfnamefont {K.~D.}\ \bibnamefont {Kihm}},
  \bibinfo {author} {\bibfnamefont {H.~g.}\ \bibnamefont {Kim}}, \bibinfo
  {author} {\bibfnamefont {G.}~\bibnamefont {Lim}}, \bibinfo {author}
  {\bibfnamefont {J.~S.}\ \bibnamefont {Park}}, \ and\ \bibinfo {author}
  {\bibfnamefont {J.~S.}\ \bibnamefont {Lee}},\ }\href {\doibase
  10.1038/srep06364} {\bibfield  {journal} {\bibinfo  {journal} {Sci.~Rep.}\ }\textbf {\bibinfo {volume} {4}},\ \bibinfo {pages} {6364}
  (\bibinfo {year} {2014})}\BibitemShut {NoStop}%
\bibitem [{\citenamefont {Golla}\ \emph {et~al.}(2013)\citenamefont {Golla},
  \citenamefont {Chattrakun}, \citenamefont {Watanabe}, \citenamefont
  {Taniguchi}, \citenamefont {LeRoy},\ and\ \citenamefont
  {Sandhu}}]{Golla2013OpticalFlakes}%
  \BibitemOpen
  \bibfield  {author} {\bibinfo {author} {\bibfnamefont {D.}~\bibnamefont
  {Golla}}, \bibinfo {author} {\bibfnamefont {K.}~\bibnamefont {Chattrakun}},
  \bibinfo {author} {\bibfnamefont {K.}~\bibnamefont {Watanabe}}, \bibinfo
  {author} {\bibfnamefont {T.}~\bibnamefont {Taniguchi}}, \bibinfo {author}
  {\bibfnamefont {B.~J.}\ \bibnamefont {LeRoy}}, \ and\ \bibinfo {author}
  {\bibfnamefont {A.}~\bibnamefont {Sandhu}},\ }\href {\doibase
  10.1063/1.4803041} {\bibfield  {journal} {\bibinfo  {journal} {Appl.~Phys.~Lett.}\ }\textbf {\bibinfo {volume} {102}},\ \bibinfo {pages}
  {161906} (\bibinfo {year} {2013})}\BibitemShut {NoStop}%
\bibitem [{\citenamefont {Ajayan}\ \emph {et~al.}(2016)\citenamefont {Ajayan},
  \citenamefont {Kim},\ and\ \citenamefont
  {Banerjee}}]{Ajayan2016Two-dimensionalMaterials}%
  \BibitemOpen
  \bibfield  {author} {\bibinfo {author} {\bibfnamefont {P.}~\bibnamefont
  {Ajayan}}, \bibinfo {author} {\bibfnamefont {P.}~\bibnamefont {Kim}}, \ and\
  \bibinfo {author} {\bibfnamefont {K.}~\bibnamefont {Banerjee}},\ }\href
  {\doibase 10.1063/PT.3.3297} {\bibfield  {journal} {\bibinfo  {journal}
  {Physics Today}\ }\textbf {\bibinfo {volume} {69}},\ \bibinfo {pages} {38}
  (\bibinfo {year} {2016})}\BibitemShut {NoStop}%
\bibitem [{\citenamefont {Blake}\ \emph {et~al.}(2007)\citenamefont {Blake},
  \citenamefont {Hill}, \citenamefont {Castro~Neto}, \citenamefont {Novoselov},
  \citenamefont {Jiang}, \citenamefont {Yang}, \citenamefont {Booth},\ and\
  \citenamefont {Geim}}]{Blake2007MakingVisible}%
  \BibitemOpen
  \bibfield  {author} {\bibinfo {author} {\bibfnamefont {P.}~\bibnamefont
  {Blake}}, \bibinfo {author} {\bibfnamefont {E.~W.}\ \bibnamefont {Hill}},
  \bibinfo {author} {\bibfnamefont {A.~H.}\ \bibnamefont {Castro~Neto}},
  \bibinfo {author} {\bibfnamefont {K.~S.}\ \bibnamefont {Novoselov}}, \bibinfo
  {author} {\bibfnamefont {D.}~\bibnamefont {Jiang}}, \bibinfo {author}
  {\bibfnamefont {R.}~\bibnamefont {Yang}}, \bibinfo {author} {\bibfnamefont
  {T.~J.}\ \bibnamefont {Booth}}, \ and\ \bibinfo {author} {\bibfnamefont
  {A.~K.}\ \bibnamefont {Geim}},\ }\href {\doibase 10.1063/1.2768624}
  {\bibfield  {journal} {\bibinfo  {journal} {Appl.~Phys.~Lett.}\
  }\textbf {\bibinfo {volume} {91}},\ \bibinfo {pages} {063124} (\bibinfo
  {year} {2007})}\BibitemShut {NoStop}%
\bibitem [{\citenamefont {Gorbachev}\ \emph {et~al.}(2011)\citenamefont
  {Gorbachev}, \citenamefont {Riaz}, \citenamefont {Nair}, \citenamefont
  {Jalil}, \citenamefont {Britnell}, \citenamefont {Belle}, \citenamefont
  {Hill}, \citenamefont {Novoselov}, \citenamefont {Watanabe}, \citenamefont
  {Taniguchi}, \citenamefont {Geim},\ and\ \citenamefont
  {Blake}}]{Gorbachev2011HuntingSignatures}%
  \BibitemOpen
  \bibfield  {author} {\bibinfo {author} {\bibfnamefont {R.~V.}\ \bibnamefont
  {Gorbachev}}, \bibinfo {author} {\bibfnamefont {I.}~\bibnamefont {Riaz}},
  \bibinfo {author} {\bibfnamefont {R.~R.}\ \bibnamefont {Nair}}, \bibinfo
  {author} {\bibfnamefont {R.}~\bibnamefont {Jalil}}, \bibinfo {author}
  {\bibfnamefont {L.}~\bibnamefont {Britnell}}, \bibinfo {author}
  {\bibfnamefont {B.~D.}\ \bibnamefont {Belle}}, \bibinfo {author}
  {\bibfnamefont {E.~W.}\ \bibnamefont {Hill}}, \bibinfo {author}
  {\bibfnamefont {K.~S.}\ \bibnamefont {Novoselov}}, \bibinfo {author}
  {\bibfnamefont {K.}~\bibnamefont {Watanabe}}, \bibinfo {author}
  {\bibfnamefont {T.}~\bibnamefont {Taniguchi}}, \bibinfo {author}
  {\bibfnamefont {A.~K.}\ \bibnamefont {Geim}}, \ and\ \bibinfo {author}
  {\bibfnamefont {P.}~\bibnamefont {Blake}},\ }\href {\doibase
  10.1002/smll.201001628} {\bibfield  {journal} {\bibinfo  {journal} {Small}\
  }\textbf {\bibinfo {volume} {7}},\ \bibinfo {pages} {465} (\bibinfo {year}
  {2011})}\BibitemShut {NoStop}%
\bibitem [{\citenamefont {Frey}\ and\ \citenamefont
  {Helmut}(2015)}]{Frey2015MeasurementsProcess}%
  \BibitemOpen
  \bibfield  {author} {\bibinfo {author} {\bibfnamefont {H.}~\bibnamefont
  {Frey}}\ and\ \bibinfo {author} {\bibfnamefont {T.}~\bibnamefont {Helmut}},\
  }in\ \href {\doibase 10.1007/978-3-642-05430-3{\_}12} {\emph {\bibinfo
  {booktitle} {Handbook of Thin-Film Technology}}}\ (\bibinfo  {publisher}
  {Springer Berlin Heidelberg},\ \bibinfo {address} {Berlin, Heidelberg},\
  \bibinfo {year} {2015})\ pp.\ \bibinfo {pages} {301--355}\BibitemShut
  {NoStop}%
\bibitem [{\citenamefont {Tolansky}(1951)}]{Tolansky1951TheInterferometry}%
  \BibitemOpen
  \bibfield  {author} {\bibinfo {author} {\bibfnamefont {S.}~\bibnamefont
  {Tolansky}},\ }\href {\doibase 10.1364/JOSA.41.000425} {\bibfield  {journal}
  {\bibinfo  {journal} {J.~Opt.~Soc.~Am.}\ }\textbf
  {\bibinfo {volume} {41}},\ \bibinfo {pages} {425} (\bibinfo {year}
  {1951})}\BibitemShut {NoStop}%
\bibitem [{\citenamefont {Raz}(1996)}]{Raz1996AInterferometer}%
  \BibitemOpen
  \bibfield  {author} {\bibinfo {author} {\bibfnamefont {E.}~\bibnamefont
  {Raz}},\ }\href {\doibase 10.1063/1.1147152} {\bibfield  {journal} {\bibinfo
  {journal} {Rev.~Sci.~Instrum.}\ }\textbf {\bibinfo {volume}
  {67}},\ \bibinfo {pages} {3416} (\bibinfo {year} {1996})}\BibitemShut
  {NoStop}%
\bibitem [{\citenamefont {O’Keefe}\ and\ \citenamefont
  {Deacon}(1988)}]{OKeefe1988}%
  \BibitemOpen
  \bibfield  {author} {\bibinfo {author} {\bibfnamefont {A.}~\bibnamefont
  {O’Keefe}}\ and\ \bibinfo {author} {\bibfnamefont {D.~A.~G.}\ \bibnamefont
  {Deacon}},\ }\href {\doibase doi:http://dx.doi.org/10.1063/1.1139895}
  {\bibfield  {journal} {\bibinfo  {journal} {Rev.~Sci.~Instrum.}\ }\textbf {\bibinfo {volume} {59}},\ \bibinfo {pages} {2544}
  (\bibinfo {year} {1988})}\BibitemShut {NoStop}%
\bibitem [{\citenamefont {Berden}\ and\ \citenamefont
  {Engeln}(2009)}]{2009CavitySpectroscopy}%
  \BibitemOpen
  \bibinfo {editor} {\bibfnamefont {G.}~\bibnamefont {Berden}}\ and\ \bibinfo
  {editor} {\bibfnamefont {R.}~\bibnamefont {Engeln}},\ eds.,\ \href {\doibase
  10.1002/9781444308259} {\emph {\bibinfo {title} {{Cavity Ring-Down
  Spectroscopy}}}}\ (\bibinfo  {publisher} {John Wiley {\&} Sons, Ltd},\
  \bibinfo {address} {Chichester, UK},\ \bibinfo {year} {2009})\BibitemShut
  {NoStop}%
\bibitem [{\citenamefont {Richardson}\ \emph {et~al.}(2009)\citenamefont
  {Richardson}, \citenamefont {Walker}, \citenamefont {Grant}, \citenamefont
  {Stoppa}, \citenamefont {Borghetti}, \citenamefont {Charbon}, \citenamefont
  {Gersbach},\ and\ \citenamefont {Henderson}}]{Richardson2009AImaging}%
  \BibitemOpen
  \bibfield  {author} {\bibinfo {author} {\bibfnamefont {J.}~\bibnamefont
  {Richardson}}, \bibinfo {author} {\bibfnamefont {R.}~\bibnamefont {Walker}},
  \bibinfo {author} {\bibfnamefont {L.}~\bibnamefont {Grant}}, \bibinfo
  {author} {\bibfnamefont {D.}~\bibnamefont {Stoppa}}, \bibinfo {author}
  {\bibfnamefont {F.}~\bibnamefont {Borghetti}}, \bibinfo {author}
  {\bibfnamefont {E.}~\bibnamefont {Charbon}}, \bibinfo {author} {\bibfnamefont
  {M.}~\bibnamefont {Gersbach}}, \ and\ \bibinfo {author} {\bibfnamefont
  {R.~K.}\ \bibnamefont {Henderson}},\ }in\ \href {\doibase
  10.1109/CICC.2009.5280890} {\emph {\bibinfo {booktitle} {2009 IEEE Custom
  Integrated Circuits Conference}}}\ (\bibinfo  {publisher} {IEEE},\ \bibinfo
  {year} {2009})\ pp.\ \bibinfo {pages} {77--80}\BibitemShut {NoStop}%
\bibitem [{\citenamefont {Gigan}\ \emph
  {et~al.}(2005{\natexlab{b}})\citenamefont {Gigan}, \citenamefont {Lopez},
  \citenamefont {Treps}, \citenamefont {Ma{\^{i}}tre},\ and\ \citenamefont
  {Fabre}}]{Gigan2005}%
  \BibitemOpen
  \bibfield  {author} {\bibinfo {author} {\bibfnamefont {S.}~\bibnamefont
  {Gigan}}, \bibinfo {author} {\bibfnamefont {L.}~\bibnamefont {Lopez}},
  \bibinfo {author} {\bibfnamefont {N.}~\bibnamefont {Treps}}, \bibinfo
  {author} {\bibfnamefont {A.}~\bibnamefont {Ma{\^{i}}tre}}, \ and\ \bibinfo
  {author} {\bibfnamefont {C.}~\bibnamefont {Fabre}},\ }\href
  {http://link.aps.org/doi/10.1103/PhysRevA.72.023804} {\bibfield  {journal}
  {\bibinfo  {journal} {Phys.~Rev.~A}\ }\textbf {\bibinfo {volume} {72}},\
  \bibinfo {pages} {23804} (\bibinfo {year} {2005}{\natexlab{b}})}\BibitemShut
  {NoStop}%
\end{thebibliography}
\end{document}